\documentclass[11pt]{article}

\usepackage{amsmath,dsfont}
\usepackage[english]{babel}
\usepackage{amsfonts}
\usepackage{amsmath,amssymb}
\usepackage{graphicx}

\makeatletter

\def\Z{\mathbb Z}
\def\R{\mathbb R}
\def\A{\mathbb A}
\def\W{\mathbb W}
\def\Y{\mathbb Y}
\def\X{\mathbb X}
\def\J{\mathbb J}
\def\P{\mathbb P}

\def\eqnarray{\stepcounter{equation}\let\@currentlabel=\theequation
\global\@eqnswtrue
\global\@eqcnt\z@\tabskip\@centering\let\\=\@eqncr
$$\halign to \displaywidth\bgroup\@eqnsel\hskip\@centering
  $\displaystyle\tabskip\z@{##}$&\global\@eqcnt\@ne
  \hfil$\displaystyle{\hbox{}##\hbox{}}$\hfil
  &\global\@eqcnt\tw@ $\displaystyle\tabskip\z@
  {##}$\hfil\tabskip\@centering&\llap{##}\tabskip\z@\cr}
\@addtoreset{equation}{section}
  \def\theequation{\thesection.\arabic{equation}}
\makeatother

\usepackage{amsmath,amssymb}
\def\beq{\begin{equation}}
\def\eeq{\end{equation}}
\def\beqa{\begin{eqnarray}}
\def\eeqa{\end{eqnarray}}
\def\barray{\begin{array}}
\def\earray{\end{array}}

\begin{document}

\title{{\bf     Fermion in a multi-kink-antikink
soliton background,\\
and exotic supersymmetry
}
}

\author{\textsf{Adri\'an Arancibia${}^{a}$,
Juan Mateos Guilarte${}^{b}$ and
Mikhail S. Plyushchay${}^{a}$}
\\
[4pt]
 {\small \textit{${}^{a}$ Departamento de F\'{\i}sica, Universidad de
Santiago de Chile, Casilla 307, Santiago 2,
Chile}}\\
{\small \textit{${}^{b}$ Departamento de F\'{\i}sica Fundamental
and IUFFyM, Universidad de Salamanca, Salamanca E-37008, Spain  }}\\
\sl{\small{E-mails: adaran.phi@gmail.com, guilarte@usal.es,
mikhail.plyushchay@usach.cl} }}
\date{}

\maketitle

\begin{abstract}
 We construct  a fermion system in a multi-kink-antikink soliton
 background,  and present in an explicit
form all its trapped configurations (bound state solutions) as
well as scattering states. This is achieved by exploiting an
exotic $N=4$ centrally extended nonlinear  supersymmetry of
completely isospectral pairs of reflectionless Schr\"odinger
systems with potentials  to be $n$-soliton solutions for the
Korteweg-de Vries equation. The obtained reflectionless Dirac
system with a position-dependent mass is shown to  possess its own
exotic nonlinear supersymmetry associated with the matrix
Lax-Novikov operator being a Darboux-dressed momentum.
 In the process, we get an algebraic
recursive representation for the multi-kink-antikink backgrounds,
and establish their relation to the the modified Korteweg-de Vries
equation. We also  indicate how the results can be related to the
physics of self-consistent condensates based on the Bogoliubov-de
Gennes equations.
\end{abstract}

\vskip.5cm\noindent

\section{Introduction and summary}

Fermion systems in soliton backgrounds describe a variety of
phenomena in particle, condensed matter and atomic physics. The
applications include, {\it inter alia}, hadron physics,
charge and fermion number
fractionalization,   conducting polymers,
superfluidity,  superconductivity,
and Bose-Einstein condensation \cite{GN}--\cite{Yef}.
The properties of such systems are
inherently related to different  aspects of
symmetries of the very diverse nature.
Much attention to investigation of
fermions in soliton backgrounds was given
in the context of supersymmetry \cite{WitOl}--\cite{BriDel}.

Classical solitons and quantum reflectionless systems
are known to be intimately related \cite{GGKM,KayMos}.
Reflectionless
potentials associated with the
soliton solutions to the Korteweg-de Vries (KdV)
equation can be constructed, particularly,  by applying
Darboux-Crum  transformations to a free
Schr\"odinger particle \cite{MatSal}.
In this picture there
appear two distinct differential operators of the even and odd
orders, which intertwine a reflectionless Hamiltonian supporting
$n$ bound states with the Schr\"odinger Hamiltonian of the same
$n$-soliton type \cite{PAM}. Any pair of $n$-soliton Schr\"odinger
systems can be  described then by an exotic non-linear
supersymmetry. This is generated \emph{not by two}, as it should
be expected for an ordinary supersymmetric pair of Hamiltonians,
but by four \emph{higher-order} differential supercharges
alongside with the two bosonic integrals having the nature of the
Lax-Novikov operators of the KdV hierarchy. Such exotic
supersymmetry was studied  by us recently in \cite{PAM}, where  we
found that its general structure, particularly the differential
order of the irreducible supercharges,
 depends essentially  on a relation
between the scattering data of the partner
Hamiltonian operators.

In this paper we show that within a family of \emph{completely
isospectral pairs} of the $n$-soliton systems, there is a peculiar
subset for which two of the four supercharges
are the \emph{ first order}
matrix differential operators, while  other
two have the differential order $2n$.
The first order supercharges are
composed from differential operators intertwining the
isospectral reflectionless partners directly.  The
supersymmetry associated with them
 is  spontaneously broken,
and the scale of the breaking is correlated with a relative shift
of soliton phases of the partner potentials. Other pair of
supercharges is constructed from the operators that intertwine the
Hamiltonians via a virtual free particle Schr\"odinger system. One
of the two nontrivial bosonic integrals, which are the Lax-Novikov
differential operators of  order $2n+1$, transmutes, in comparison
with a general case of $n$-soliton paired  systems, into a central
charge of the nonlinear superalgebra. The condition of
commutativity of the central charge with any of the two  first
order supercharges can be interpreted as a stationary equation  of
the hierarchy of the modified Korteweg-de Vries (mKdV) system
represented according to
 the Zakharov-Shabat -- Ablowitz-Kaup-Newell-Segur
 (ZS-AKNS) \cite{ZS,AKNS}
 $2\times 2$  matrix scheme.
 The second nontrivial bosonic integral
 generates a kind of  rotation between
 the two types of supercharges.

 A remarkable possibility for alternative interpretation of one of
 the two first order supercharges as a Hamiltonian
 of a Dirac particle with a position-dependent mass
 provides us then with a fermion system in a
 multi-kink-antikink soliton background.
 All the scattering and bound states
 (trapped configurations)
 of the fermion system
 are constructed by Darboux dressing of
 the free massive Dirac  particle.
The obtained reflectionless Dirac system is shown to
 possess its own exotic nonlinear supersymmetry
 that effectively encodes its spectral peculiarities.
In the process, we get a recursive representation
for the multi-kink-antikink backgrounds.
We also  indicate how to relate
the results to the physics of self-consistent
condensates based on  the
Bogoliubov-de Gennes (BdG) equations.
In this context, the multi-kink-antikink backgrounds
we construct  and study correspond to
a generalization of the fermion-antifermion bound states
solutions of Dashen, Hasslacher and Neveu
\cite{DHN} for the Gross-Neveu model \cite{GN}.
In the last years, investigation of self-consistent solutions
to the Gross-Neveu  model and physics related to them experiences
a renovation of interest
\cite{Feinberg,ThiesM,SchonTh,FeinHill,BasD,PAN,TakNit,DunneThies}.

The paper is organized as follows. In the next section we
summarize shortly the general properties of the Schr\"odinger
$n$-soliton potentials constructed by inverse scattering method,
and their relation to the KdV evolution equation and to the
stationary KdV hierarchy. Then we discuss a  construction of the
corresponding reflectionless $n$-soliton systems from a free
Schr\"odinger particle by means of the Darboux-Crum
transformations, and show a relation of them to the non-linear
Schr\"odinger equation. We also obtain a recursive
representation for the multi-soliton potentials,  and describe
briefly how the exotic supersymmetric structure of a general form
emerges in the extended quantum systems composed from the  pairs
of reflectionless $n$-soliton Schr\"odinger Hamiltonians. In
Section 3 we prove that for any value of $n$, there is a
very special
$(2n+1)$-parametric $2\times 2$ matrix quantum system
given by a pair of completely isospectral $n$-soliton
Schr\"odinger partners
 intertwined by the first order differential operators.
We also present there the
explicit form of the superalgebra of the
corresponding exotic $N=4$ centrally extended
nonlinear  supersymmetry.
We reinterpret
the obtained special class
of supersymmetric systems in Section 4  by
considering one of its two first order supercharges as a Dirac
Hamiltonian. The obtained fermion system in a multi-kink-antikink
background is associated
then with the mKdV evolution system
presented in the ZS-AKNS
$2\times 2$ matrix scheme. In Section 5 the reflectionless
fermion  system is treated as a Darboux-dressed
form of the free massive Dirac
particle, and  its own exotic nonlinear supersymmetry is
identified. Section 6 is devoted to the concluding comments,
where we discuss briefly  some further interesting developments
and applications of the results. We indicate, particularly,
how they can be  related to the physics of self-consistent
condensates with both zero and nonzero values
of a topological charge.  In
two Appendices we summarize shortly
some aspects of the Dabroux
and Miura transformations, which are used  in the main text.

\section{Reflectionless Schr\"odinger potentials
and exotic supersymmetry}

We review here briefly some properties of the soliton solutions to
the KdV equation, and identify the exotic nonlinear supersymmetric
structure of the extended systems composed from the reflectionless
pairs of $n$-soliton  Schr\"odinger Hamiltonians. In the process
we observe a relation of the bound state eigenvalue problem for
the  $n$-soliton potential with the coupled system of non-linear
Schr\"odinger equations, and obtain a recursive representation for
multi-soliton potentials.

\subsection{Reflectionless potentials and the KdV }

There exists a variety of  possible ways to construct
reflectionless quantum mechanical systems.  This can be done,
particularly,  by the inverse scattering method
\cite{KayMos,GGKM}, by B\"acklund \cite{WalEsta,DraJoh}, and by
Darboux-Crum \cite{MatSal} transformations. In the inverse
scattering method, a reflectionless potential supporting $n$ bound
states can be presented in a form \cite{KayMos,GGKM}
\begin{equation}\label{U}
    U_n(x)=-2\frac{d}{d x}K_n(x,x),\qquad
    K_n(x,x)=\frac{d}{dx}\ln\,[\det   \mathcal{K}]\,.
\end{equation}
Here $  \mathcal{K}$ is the  $n\times n$  matrix with elements
\begin{equation}\label{ANN}
    \mathcal{K}_{ij}=
    \delta_{ij}+\frac{\beta_i\beta_j}{\kappa_i+\kappa_j}\,
    e^{-(\kappa_i+\kappa_j)x}
\end{equation}
given in terms of $2n$ real parameters $\kappa_j$ and $\beta_j$,
$j=1,\ldots,n$, $\kappa_n>\kappa_{n-1}>\ldots\kappa_1>0$,
$\beta_j>0$. Parameters $\kappa_j$ correspond to the energies of
the bound states, $E_j=-\kappa_j^2$, and $\beta_j$ are associated
with their normalization constants. Reflectionless potential
$U_n(x)$ satisfies an ordinary nonlinear differential equation of
order $2n+1$, that is a so-called Novikov equation, or a
stationary equation of the KdV hierarchy \cite{SPN}.

Introducing a dependence of $\beta_j$ on an evolution parameter
$t$ in the form  $\beta_j(t)=\beta_j(0)e^{4\kappa_j^3t}$, we
obtain a function $U_n(x,t)$,  which describes an $n$-soliton
solution to the KdV equation \cite{GGKM}
$$
u_t-6uu_x+u_{xxx}=0\,,
$$
where $u_t=\frac{\partial}{\partial t}u$,
$u_x=\frac{\partial}{\partial x}u$. For large positive and
negative values of $t$, the  $U_n(x,t)$ decouples into a linear
sum of the $n$ one-soliton solutions of the amplitudes
$2\kappa_j^2$, which move to the right at the speeds
$v_j=4\kappa_j^2$,
\begin{equation}\label{U1}
    U_{n}(x,t)=-\sum_{j=1}^n 2\kappa_j^2{\rm sech}\,^2
    \kappa_j(x-4\kappa^2_jt\pm x^\pm_{0j})\,\quad
    {\rm as}\quad t\rightarrow\pm \infty\,.
\end{equation}
The  phases, or centers  $x^\pm_{0j}$ of solitons are expressed in
terms of the $\beta_j(0)$ and scaling parameters $\kappa_j$. At
finite values of  $t$, the $U_n(x,t)$ describes a nonlinear
interaction of $n$ solitons. As a result of the soliton
scattering, the phases  suffer certain displacements,
$x^+_{0j}-x^-_{0j}=\Delta x_{0j}(\kappa)$, which depend only on
the scaling parameters \cite{NovZak}.

A choice of $\beta_j(t)=\beta_j(0)\exp (P_{2\ell+1}(\kappa_j) t)$
instead of $\beta_j(t)=\beta_j(0)e^{4\kappa_j^3t}$, where
$P_{2\ell+1}(\kappa)$ is an odd polynomial
$P_{2\ell+1}(\kappa)=a_{\ell}\kappa^{2\ell+1}+a_{\ell-1}
\kappa^{2\ell-1}+\ldots +a_1\kappa^3+a_0\kappa$ given in terms of
a set of constants  $a_0,\ldots, a_{\ell}$, generates an
$n$-soliton potential which will evolve in time in accordance with
some  equation of the KdV hierarchy.

\subsection{Darboux-Crum transformations, reflectionless potentials,
and nonlinear Schr\"odinger
equation}

Another representation  of the soliton systems, which is based on
the Darboux transformations, is more convenient for the
supersymmetric structure we are going to study. The Schr\"odinger
Hamiltonian $H_n=H_0+U_n(x)$ of a reflectionless system with $n$
bound states can be obtained by applying the Darboux-Crum
transformation, which is a composition of $n$ Darboux
transformations,  to a free particle described by
$H_0=-\frac{d^2}{dx^2}$. A reflectionless potential in this case
is represented as
\begin{equation}\label{UNDC}
    U_n(x)=-2\frac{d^2}{dx^2}\ln  \W_n(x)\
\end{equation}
in terms of the Wronskian $\W_n(x)=\W(\psi_1,\ldots, \psi_n)$,
$\W(f_1,\ldots,f_n)=\det W_{ij}$,
$W_{ij}=\frac{d^{i-1}}{dx^{i-1}}f_j$, which is constructed from
\emph{non-physical}, exponentially divergent at infinity
eigenfunctions $\psi_j$ of the free particle Hamiltonian,
$H_0\psi_j=-\kappa_j^2\psi_j$,
\begin{equation}\label{psij0}
    \psi_j(x;\kappa_j,\tau_j)=
    \left\{\begin{matrix}\cosh \kappa_j(x+\tau_j),&
    j={\rm odd}\\ \sinh
    \kappa_j(x+\tau_j),&j={\rm even}\end{matrix}\,\,.
    \right.
\end{equation}
The scaling
parameters $\kappa_j$,
$0<\kappa_1<\kappa_2<...<\kappa_{j-1}<\kappa_n$,
 are the same here as in (\ref{ANN}),
while
the translation  parameters $\tau_j$, $j=1,\ldots,n$, may take
arbitrary real values, and can be related with the
parameters $\beta_j$ in representation (\ref{U}), (\ref{ANN}).
The subsets of wave functions (\ref{psij0})
with even and odd values of index $j$
can be transformed mutually into each other
by a complex shift of the translation parameters, $\cosh
\kappa_j(x+\tau_j+i\frac{\pi}{2\kappa_j})=i\sinh
\kappa_j(x+\tau_j)$, or by a differentiation. A specific choice of
the free particle Hamiltonian eigenstates in
(\ref{psij0})
guarantees that the Wronskian $\W_n(x)$
is a nodeless function
that generates a nontrivial,
$2n$-parametric non-singular potential
(\ref{UNDC}) \cite{PAM},
$
    U_n=U_n(x;\kappa_1,\ldots,\kappa_n,\tau_1,\ldots,\tau_n).
$
The Wronskian here can be related to the determinant
in representation (\ref{U}), (\ref{ANN}),
$\W_n(x)=Ce^{\rho x}\det \mathcal{K}$, where
$C=C(\kappa,\tau)$ and $\rho=\rho(\kappa,\tau)$ are some
constants.

According to the Darboux-Crum construction, the eigenstates
$\psi[n,\lambda]$ of the Schr\"odinger operator $H_n$,
$H_n\psi[n,\lambda]= \lambda\psi[n,\lambda]$, are obtained from
the free particle eigenfunctions $\psi[0,\lambda]$,
$H_0\psi[0,\lambda]= \lambda\psi[0,\lambda]$,
\begin{equation}\label{psiNlam}
    \psi[n;\lambda]=\frac{\W(\psi_1,\ldots,\psi_n,
    \psi[0;\lambda])}{\W(\psi_1,\ldots,\psi_n)}\,\,.
\end{equation}
Un-normalized physical bound states $\psi[n,-\kappa_j^2]$,
$j=1,\ldots,n$, are constructed, particularly, from the non-physical
free particle eigenfunctions
\begin{equation}\label{psij0til}
    \psi[0,-\kappa_j^2](x)\equiv{\psi}'_j(x;\kappa_j,\tau_j)=
    \left\{\begin{matrix}\sinh \kappa_j(x+\tau_j),&
    j={\rm odd}\\ \cosh
    \kappa_j(x+\tau_j),& j={\rm even}\end{matrix}\,.
    \right.
\end{equation}
Functions (\ref{psij0til}) form a set complementary to
(\ref{psij0}), $H_0 {\psi}'_j=-\kappa_j^2{\psi}'_j$. As it was
noted, the set (\ref{psij0til}) can be related to (\ref{psij0}) by
a simple complex shift of translation parameters, or by a
differentiation,
\begin{equation}\label{psitdif}
    {\psi}'_j(x;\kappa_j,\tau_j)=\frac{1}{\kappa_j}
\frac{d}{dx}\psi_j(x;\kappa_j,\tau_j)\,.
\end{equation}

\vskip0.1cm

Relation
 (\ref{psiNlam})  can be
presented in an equivalent form
\begin{equation}\label{AApsi}
    \psi[n;\lambda]=\A_n\psi[0;\lambda]\,,\qquad
    \A_n=A_nA_{n-1}\ldots
    A_1\,,
\end{equation}
which will play a key role in the further analysis.
Here the first order differential operators
$A_j$ are defined recursively in terms of the functions
(\ref{psij0}),
\begin{eqnarray}\label{A1def}
    &A_1=\psi_1\frac{d}{dx}\frac{1}{\psi_1}=
    \frac{d}{dx}-(\ln \psi_1)_x,&\\
    &A_j=(\A_{j-1}\psi_j)\frac{d}{dx}\frac{1}{(\A_{j-1}
    \psi_j)}=\frac{d}{dx}-(\ln (\A_{j-1}
    \psi_j))_x\,,\quad
    j=2,\ldots\,.&\label{Ajdef}
\end{eqnarray}
Indeed, the equivalence  of (\ref{AApsi}) to (\ref{psiNlam}) for
$n=1,2$ is checked directly. Assuming that
\begin{equation}\label{WNAN}
     \A_n\psi[0;\lambda]= \frac{\W(\psi_1,\ldots,\psi_n,
    \psi[0;\lambda])}{\W(\psi_1,\ldots,\psi_n)}
\end{equation}
is valid for $n>2$,  Eqs.  (\ref{Ajdef}) and (\ref{WNAN}) give
\begin{equation}\label{ANproof}
    \A_{n+1}\psi[0;\lambda]=
    A_{n+1}\left(\A_n \psi[0;\lambda]\right)=
    (\A_n\psi_{n+1})\frac{d}{dx}
    \left(\frac{1}{(\A_n\psi_{n+1})}
    \A_n \psi[0;\lambda]\right)\,,
\end{equation}
and
\begin{eqnarray}
    \A_{n+1}\psi[0;\lambda]
    &=&\frac{\W(1,\ldots,n,n+1)}{\W(1,\ldots,n)}
    \left(\frac{\W(1,\ldots,n)}{\W(1,\ldots,n,n+1)}
    \frac{\W(1,\ldots,n,0)}{\W(1,\ldots,n)}\right)_x\nonumber\\
    &=&
    \frac{\W(\W(1,\ldots,n,n+1),\W(1,\ldots,n,0))}{
    \W(1,\ldots,n)\W(1,\ldots,n,n+1)},\label{WN+1}
\end{eqnarray}
where $\W(1,\ldots,n,n+1)=\W(\psi_1,\ldots,\psi_{n+1})$,
$\W(1,\ldots,n,0)=\W(\psi_1,\ldots,\psi_n,\psi[0;\lambda])$. The
Wronskian identity
\begin{equation}\label{Widen}
    \W(f_1,\ldots,f_n,g,h)\W(f_1,
    \ldots,f_n)=\W(\W(f_1,\ldots,f_n,g),
    \W(f_1,
    \ldots,f_n,h))\,,
\end{equation}
which is true for any choice of the functions  $f_1,\ldots, f_n$,
$g$ and $h$ \cite{Crum}, allows us to represent the fraction
(\ref{WN+1}) in the form of the right hand side of (\ref{WNAN})
with $n$ changed for $n+1$. This proves the equivalence
 of (\ref{AApsi}) to (\ref{psiNlam}) by induction.

Definition (\ref{Ajdef}) and relation (\ref{WNAN}) provide also
the  following alternative representation for the operator $A_n$,
\begin{equation}\label{Andx}
    A_n=\frac{d}{dx}-
    \left(\ln \A_{n-1}\psi_n\right)_x
    =\frac{d}{dx}-\left(\ln \frac{\W_n}{\W_{n-1}}\right)_x
    \equiv\frac{d}{dx}+\mathcal{W}_n\,,
\end{equation}
where
\begin{equation}\label{omn}
    \mathcal{W}_n=
    \Omega_n-\Omega_{n-1}\,,\qquad
    \Omega_n=-(\ln \W_n)_x\,.
\end{equation}
Then  (\ref{omn}) together with Eq. (\ref{UNDC})
 gives one more useful representation for the
$n$-soliton potential,
\begin{equation}\label{omV}
    U_n=2\Omega_{nx}\,.
\end{equation}
Having in mind this relation,  we call $\Omega_n$ a pre-potential
of the $n$-soliton system. Coherently with Eqs. (\ref{WNAN}) and
(\ref{A1def}), in (\ref{Andx}) and (\ref{omn}) we assume $\W_0=1$,
 $\Omega_0=0$, $V_0=0$, and have
 $\W_1=\cosh\kappa_1(x+\tau_1)$,
 \begin{equation}\label{V1sol}
     \Omega_1=-\kappa_1\tanh\kappa_1(x+\tau_1),\qquad
    U_1=-\frac{2\kappa_1^2}{\cosh^2\kappa_1(x+\tau_1)}\,.
\end{equation}

As follows from (\ref{Ajdef}), the first order differential
operator $A_j$ annihilates a nodeless non-physical eigenfunction
$\A_{j-1}\psi_j$ of $H_{j-1}$ of eigenvalue $-\kappa_j^2$. On the
other hand, $A_j^\dagger$ annihilates a function
$1/(\A_{j-1}\psi_j)$, which is a physical bound (ground) state of
$H_j$ of energy $-\kappa_j^2$. This means that $U_n$ and $U_{n-1}$
are related by the Darboux transformation, see Appendix A.
Explicitly, we have the relations
\begin{equation}\label{VVW}
    U_n=\mathcal{W}_n^2+\mathcal{W}_{nx} -\kappa_n^2\, ,
    \qquad
    U_{n-1}=\mathcal{W}_n^2-\mathcal{W}_{n x} -\kappa_n^2\, ,
\end{equation}
\begin{equation}\label{AAH}
    A_nA_n^\dagger=H_n+\kappa_n^2\,,\qquad
      A_n^\dagger A_n=H_{n-1}+\kappa_n^2\,.
\end{equation}
In correspondence with (\ref{AAH}),
the first order Darboux generators $A_n$
and $A_n^\dagger$ intertwine
the $n$- and $(n-1)$-soliton systems,
$$
    A_nH_{n-1}=H_nA_n,\qquad
    A_n^\dagger H_n=H_{n-1}A_n^\dagger,
    $$
and relate their eigenstates,
$$
    \psi[n;\lambda]=A_n\psi[n-1;\lambda],\qquad
    A_n^\dagger\psi[n;\lambda]=
    (\lambda +\kappa_n^2)\psi[n-1;\lambda],
$$
cf. (\ref{AApsi}).
The order $n$ differential operators $\A_n$ and $\A_n^\dagger$
intertwine,  on the other hand, $H_n$
with a free particle Hamiltonian $H_0$,
    \begin{equation}\label{H0A}
\A_nH_{0}=H_n\A_n,\qquad
 \A_n^\dagger H_n=H_{0}\A_n^\dagger\, .
\end{equation}

As follows from (\ref{AApsi}),  the free particle's plane wave
states $e^{ikx}$  are mapped into the eigenfunctions of $H_n$ of
the form $\psi_n(x,k)=P_n(x,k)e^{ikx}$, where $P_n$ is a
polynomial of order $n$ in  $k$, $H_n\psi_n(x,k)=k^2\psi_n(x,k)$.
This means that $U_n(x)$ is a Bargmann-Kay-Moses  reflectionless
potential \cite{KayMos}, for which the transmission coefficient
can be easily computed. For functions (\ref{psij0}) we have
$\psi_j(x)\sim e^{\pm \kappa_j(x+\tau_j)}$ as $x\rightarrow
\pm\infty$. Then we find that $A_j\rightarrow \frac{d}{dx}\pm
\kappa_j$  as $x\rightarrow \mp \infty$, and in these limits
$P_n\rightarrow P_{n\mp}= \prod_{j=1}^n(ik\pm \kappa_j)$. For the
transmission amplitude $t(k)=P_{n+}/P_{n-}$ this gives
\begin{equation}\label{trans}
    t(k)=\prod_{j=1}^{n}
    \left(\frac{k+i\kappa_j}{k-i\kappa_j}\right)\,.
\end{equation}

A class of reflectionless systems
we consider turns out also to be related naturally to another
completely integrable system, namely, to the
nonlinear Schr\"odinger equation.

To see this, we first show that the reflectionless potential
$U_n(x)$ can be presented in the form
\begin{equation}\label{UNKayMos}
    U_n(x)=-4\sum_{j=1}^{n}\kappa_j
    \hat{\psi}^2_{n,j}(x)
\end{equation}
in terms of  the normalized  bound states of the Hamiltonian
$H_n$,
\begin{equation}\label{psihat}
    \hat{\psi}_{n,j}(x)=
    \mathcal{N}^{-1}_j\psi[n,-\kappa_j^2](x),\qquad
    \mathcal{N}^2_j=2\kappa_j\prod_{\ell=1,\, \ell\neq j}^n
    |\kappa_\ell^2-\kappa_j^2|\,,\qquad
    \int_{-\infty}^{+\infty}\hat{\psi}_{n,j}^2(x)dx=1\,,
\end{equation}
where it is assumed that at $n=1$
the product in expression
for  $\mathcal{N}^{2}_1$ is reduced to $1$.
Using relation $\frac{d}{dx}\W_n= \sum_{j=1}^{n}
\W(\psi_1,\ldots,\frac{d\psi_j}{dx},\ldots,\psi_n)$, we can
rewrite
 Eq.  (\ref{UNDC}) in a form
$
    U_n(x)=-2\sum_{j=1}^n{\W\left(\W_n,
    \W(\psi_1,\ldots,\frac{d\psi_j}{dx},\ldots,\psi_n)
    \right)}/{\W_n^2}.
$
The Wronskian identity
(\ref{Widen})
 allows us to represent the potential equivalently as
\begin{equation}\label{UNxj+}
    U_n(x)=-2\sum_{j=1}^n\frac{
    \W(\psi_1,\ldots,\psi_j,\frac{d\psi_j}{dx},\ldots,\psi_n)
    \W(\psi_1,\ldots,\psi_{j-1},\psi_{j+1},\ldots,
    \psi_n)}{\W_n^2}\,.
\end{equation}
A relation
\begin{equation}\label{Wusef}
    \W(\psi_1,\ldots,\psi_j,\frac{d\psi_j}{dx},
    \ldots,\psi_n)=\frac{1}{2}
    \kappa_j\mathcal{N}_j^2 \W(\psi_1,\ldots,\psi_{j-1},
    \psi_{j+1},\ldots,
    \psi_n),
\end{equation}
where $\mathcal{N}_j^2$ is defined in (\ref{psihat}), follows from
basic identities of determinants. Using this last relation
together with Eqs. (\ref{psiNlam}), (\ref{psij0til}) and
(\ref{psitdif}), we rewrite (\ref{UNxj+}) in terms of
un-normalized bound states of $H_n$,
\begin{equation}\label{UnNnon}
    U_n(x)=-4\sum_{j=1}^n \kappa_j\mathcal{N}_j^{-2}
    \psi^2[n,-\kappa_j^2](x)\,.
\end{equation}
Employing once more the identity (\ref{Widen}) we get
\begin{eqnarray}
    &&\frac{d}{dx}\left(\frac{\W(\psi_1,
    \ldots,\frac{d\psi_j}{dx},\ldots,\psi_n)}
    {\W_n}\right)=\nonumber\\
    &&\frac{\W(\psi_1,\ldots,\psi_{j-1},\psi_{j+1},\ldots,
    \psi_n,\psi_j,\frac{d\psi_j}{dx})
    \W(\psi_1,\ldots,\psi_{j-1},\psi_{j+1},\ldots,
    \psi_n)}{\W_n^2}\,.
\end{eqnarray}
 Eq. (\ref{Wusef}) gives us then
$
\frac{d}{dx}({\W(\psi_1,\ldots,\frac{d\psi_j}{dx},\ldots,\psi_n)}
    /{\W_n})=2\kappa_j\mathcal{N}_j^{-2}\psi^2[n,-\kappa_j^2](x)$.
Integrating this equality from $-\infty$ to $+\infty$, and using
relations
    $\lim_{x\rightarrow\pm\infty}
    {\W(\psi_1,\ldots,\frac{d\psi_j}{dx},\ldots,\psi_n)}/
    {\W_n}=\pm\kappa_j\,,$
we reproduce (\ref{psihat}), and present (\ref{UnNnon})   in  the
form (\ref{UNKayMos}). \vskip0.1cm

Because of relation  (\ref{UNKayMos}), the equations
$H_n\hat{\psi}_{n,j}=-\kappa_j^2\hat{\psi}_{n,j}$ for $n$
normalized bound states can be presented in a form of the system
of $n$ coupled nonlinear ordinary differential equations
\begin{equation}\label{psijN}
    -\hat{\psi}_{n,jxx}-4\sum_{i=1}^n\kappa_i\hat{\psi}_{n,i}^2
    \hat{\psi}_{n,j} +\kappa_j^2\hat{\psi}_{n,j}=0\,.
\end{equation}
Introduce an evolution parameter $t$, and define
$q_j(x,t)=\exp(i\kappa_j^2t)\hat{\psi}_{n,j}(x)$. Then we find
that these functions satisfy a system of $n$ coupled nonlinear
Schr\"odinger equations,
\begin{equation}\label{psijN+}
    iq_{jt}=-q_{jxx}-4\sum_{i=1}^n
    \kappa_i\vert q_i\vert^2q_j\,.
\end{equation}
In the simplest case $n=1$, this reduces to a
focusing case of the
nonlinear Schr\"odinger equation,
\begin{equation}\label{selffocSch}
    iq_t+q_{xx}+4\kappa\vert q\vert^2q=0\,.
\end{equation}

So,  $n$ bound state solutions to the linear time-dependent
quantum  Schr\"odinger equation
for  reflectionless time-independent $n$-soliton potential
provide  also a solution to the system of $n$
coupled nonlinear Schr\"odinger
equations.

\subsection{Recursions for $n$-soliton pre-potentials  and potentials}

Here we obtain a recursion representation for $n$-soliton potentials
of a general form. This will allow us in what follows to get
also  a recursion representation for multi-kink-antikink
backgrounds, which are reflectionless Dirac potentials.

Let us take a sum of
two relations in (\ref{VVW}) with making use
of (\ref{omn}),
\begin{equation}\label{UnUn-1}
    U_n+U_{n-1}=2(\Omega_n-\Omega_{n-1})^2-2\kappa_n^2\,.
\end{equation}
Changing  in (\ref{UnUn-1}) $n$ for $j$, we multiply both sides of
the equality by $(-1)^{n-j}$, and sum up from $j=1$ to $j=n$. As
a  result we obtain
\begin{equation}\label{Vnsum}
    \frac{1}{2}\,U_n=\Omega_n^2+\sum_{j=1}^{n}
    (-1)^{n-j+1}\left(
    2\Omega_j\Omega_{j-1}+\kappa_j^2\right)\,.
\end{equation}
Assume now that the chain of reflectionless potential $U_j$ with
$j=1\ldots, n$ is constructed by using  the same chain  of states
(\ref{psij0}) in which, however, the last two states, $\psi_{n-1}$
and $\psi_{n}$, are permuted. In such a way we get a chain of
functions
 $\Omega_1(1),\ldots$, $\Omega_{n-2}(1,\ldots,n-2)$,
$\Omega_{n-1}(1,\ldots,n-2,n)$, $\Omega_n(1,\ldots,n-2,n,n-1)$.
Since  $\W(1,\ldots,n-2,n,n-1)=-\W(1,\ldots,n-2,n-1,n)$,
we have $\Omega_n(1,\ldots,n-2,n,n-1)=\Omega_n(1,\ldots,n-2,n-1,n)$,
and  in the indicated chain of pre-potentials only the penultimate
term $\Omega_{n-1}(1,\ldots,n-2,n)$ is different from the corresponding term
$\Omega_{n-1}(1,\ldots,n-2,n-1)$ in the  initial,
non permuted chain.
The same is valid for corresponding
chain of potentials by virtue  of relation (\ref{omV}).
 Notice that  ${\Omega}_{n-1}^\sharp\equiv
 \Omega_{n-1}(1,\ldots,n-2,n)$ and
${U}_{n-1}^\sharp\equiv U_{n-1}(1,\ldots,n-2,n)$ are
singular functions of $x\in\R$. Particularly,
\begin{equation}\label{til1O}
    \Omega_1^\sharp =\Omega_1(2)=-\kappa_2\coth \kappa_2(x+\tau_2)\,,
    \qquad
    U_1^\sharp={U}_1(2)=\frac{2\kappa_2^2}{\sinh^2\kappa_2(x+\tau_2)}\,.
\end{equation}
Let us write  the analog of relation (\ref{Vnsum}) assuming
that we construct $U_n$ via the described chain
with permuted two last states,
\begin{equation}\label{Vnsum+}
    \frac{1}{2}\,U_n=\Omega_n^2+\sum_{j=1}^{n-2}
    (-1)^{n-j+1}\left(
    2\Omega_j\Omega_{j-1}+\kappa_j^2\right)
    +(2{\Omega}_{n-1}^\sharp\Omega_{n-2}+\kappa_n^2)-
    (2\Omega_{n}{\Omega}_{n-1}^\sharp
    +\kappa_{n-1}^2)\,.
\end{equation}
Subtracting (\ref{Vnsum+}) from (\ref{Vnsum}), we get the equality
    $2\Omega_n({\Omega}_{n-1}^\sharp-\Omega_{n-1})
    +2\Omega_{n-2}(\Omega_{n-1}-{\Omega}_{n-1}^\sharp)
    +2(\kappa_{n-1}^2-\kappa_n^2)=0,$
which gives a recursive relation for the pre-potentials  $\Omega_n$,
\begin{equation}\label{Omegarec}
    \Omega_n=\Omega_{n-2}+\frac{
    \kappa_{n-1}^2-\kappa_n^2}{\Omega_{n-1}-
    {\Omega}_{n-1}^\sharp}\,,\qquad
    n\geq 2\,.
\end{equation}
 Eq.  (\ref{Omegarec}) for the first two cases
 $n=2,3$ gives
\begin{equation}\label{Om123}
   \Omega_2= \Omega_2(1,2)=\frac{\kappa_1^2-\kappa_2^2}{\Omega_1(1)-
    \Omega_1(2)}\,,
    \quad
   \Omega_{3}= \Omega_{3}(1,2,3)=\Omega_1(1)
    +\frac{\kappa_2^2-\kappa_3^2}
    {\Omega_2(1,2)-\Omega_2(1,3)}\,,
\end{equation}
and corresponding singular pre-potentials are obtained from these
by changing the last arguments, $\Omega_2^\sharp=\Omega_2(1,3)$,
$\Omega_3^\sharp=\Omega_3(1,2,4)$. Reflectionless n-soliton
potential $U_n$ with $n\geq 2$ can  be calculated now recursively,
by employing Eqs. (\ref{omV}), (\ref{V1sol}) and (\ref{Omegarec}),
\begin{equation}\label{OmnU}
    U_n=U_{n-2}+2\frac{d}{dx}\left(\frac{\kappa_{n-1}^2-
    \kappa_n^2}{\Omega_{n-1}-
    {\Omega}_{n-1}^\sharp}\right)\,,\quad
    n=2,\dots.
\end{equation}
Particularly,  for $n=2$ Eqs. (\ref{OmnU}), (\ref{V1sol}) and (\ref{til1O})
give
\begin{equation}\label{U2}
    U_2=-2(\kappa_2^2-\kappa_1^2)
    \left(\frac{\kappa_1^2}{\cosh^2\chi_1}
    +\frac{\kappa_2^2}{\sinh^2\chi_2}\right)
    (\kappa_2\coth\chi_2-
    \kappa_1\tanh \chi_1)^{-2}\,,
\end{equation}
where $\chi_j=\kappa_j(x+\tau_j),$
    $j=1,2$.

Relation (\ref{OmnU}) together with (\ref{Omegarec}) corresponds
to the recursive representation of $n$-soliton solutions of the
KdV equation obtained by Wahlquist and  Estabrook by employing
B\"acklund transformations \cite{WalEsta,DraJoh}.

\subsection{Exotic supersymmetry of reflectionless
$n$-soliton pairs}

In this subsection we describe shortly  an exotic $N=4$ supersymmetric
 structure
 that appears in the pairs of $n$-soliton Schr\"odinger
systems of the most general form \cite{PAM}, and observe the
peculiarity of the case of completely isospectral soliton partners.
 These results will be used then in next sections to identify within the
 family of isospectral $n$-soliton pairs a very special subfamily
 related to reflectionless Dirac systems, which
 correspond to a fermion in a multi-kink-antikink soliton background.

Let us consider two reflectionless systems $H_n$ and $\tilde{H}_n$
constructed by using  two sets of the parameters,
$(\kappa_1,\ldots,\kappa_n$, $\tau_1,\ldots,\tau_n)$ and
$(\tilde{\kappa}_1,\ldots,\tilde{\kappa}_n$, $\tilde{\tau}_1,
\ldots,\tilde{\tau}_n)$. Each of these two Hamiltonians can be
related to the free particle system $H_0$ by  the corresponding
intertwining operators of order $n$, $\A_n$ and $\tilde{\A}_n$,
and by the conjugate operators $\A_n^\dagger$ and
$\tilde{\A}_n^\dagger$.  Relations (\ref{H0A}) and similar
relations for $\tilde{H}_n$ together with the observation that
$\frac{d}{dx}$ is the  integral of the free particle allow us to
construct the operators which intertwine  the $n$-soliton
reflectionless systems $H_n$ and $\tilde{H}_n$,
\begin{equation}\label{YXinter}
\Y_n=\A_n\tilde{\A}_n^\dagger\,,\qquad
\X_n=\A_n \frac{d}{dx} \tilde{\A}_n^\dagger\,,
\end{equation}
\begin{equation}\label{JJH}
\J_n\tilde{H}_n=H_n\J_n\,,\qquad
\J_n^\dagger H_n=\tilde{H}_n\J_n^\dagger, \quad
\text{where}\quad
\J_n=\Y_n\,,\,\X_n\,.
\end{equation}
Operator $\Y_n$  is the differential operator
of the even order $2n$, while  $\X_n$
is  the differential operator of the odd order $2n+1$.
On the other hand,  differential operators of order
$2n+1$,
\begin{equation}\label{defZ}
    \Z_n=\A_n \frac{d}{dx}\A_n^\dagger\,,\qquad
    \tilde{\Z}_n=\tilde{\A}_n \frac{d}{dx}
    \tilde{\A}_n^\dagger\,,
\end{equation}
being the Darboux-dressed forms  of the free
 particle integral $\frac{d}{dx}$, are the integrals
 for $H_n$ and $\tilde{H}_n$,
\begin{equation}\label{ZZH}
[\Z_n,H_n]=0,\qquad
[\tilde{\Z}_n,\tilde{H}_n]=0.
\end{equation}

Operator $\Z_n$ can be presented in  a form
$\Z_n=(-1)^n\frac{d^{2n+1}}{dx^{2n+1}}+ \sum_{j=1}^{2n}
a_{2n-j}(x)\frac{d^{2n-j}}{dx^{2n-j}}$, where coefficients
$a_{2n-j}(x)$ are some functions of the potential $U_n$ and its
derivatives $U_{nx},\ldots,$ $\frac{d^{2n-1}}{dx^{2n-1}}U_n$. The
relation of commutativity of $\Z_n$ and $H_n$,  $[\Z_n,H_n]=0$, is
the Novikov equation, or, equivalently, a stationary higher
equation of the KdV hierarchy, that defines an algebro-geometric
potential $U_n(x)$ \cite{NovZak,Belo}. In correspondence with the
Burchnall-Chaundy theorem \cite{BC}, commuting differential
operators $\Z_n$ and $H_n$ of the mutually prime orders $2n+1$ and
$2$ satisfy identically a relation $\Z_n^2=P_{2n+1}(H_n)$, where
$P_{2n+1}(H_n)=H_n\prod_{j=1}^n(H_n+\kappa_j^2)^2$ is a degenerate
spectral polynomial of the $n$-soliton system \cite{PAM}. In
correspondence with this relation, integral $\Z_n$ annihilates all
the singlet physical states, which are bound states of energies
$E_j=-\kappa_j^2$, $j=1,\ldots,n$, and the state $\psi[n;0]=\A_n
1$ of zero energy being the lowest state of the continuous part of
the spectrum, cf. Eq. (\ref{AApsi}). Other $n$ states annihilated
by $\Z_n$ are the non-physical eigenstates of $H_n$ of energies
$E_j=-\kappa_j^2$, which can be related to the corresponding bound
states by equation of the form (\ref{secondsol}).

In the simplest case $n=1$,  the pre-prepotential and potential
are given by Eq. (\ref{V1sol}), and we have
$
\Z_1=\frac{1}{4}\mathcal{Z}_1+\kappa_1^2\mathcal{Z}_0,
$
where $\mathcal{Z}_0=\frac{d}{dx}$ and
$\mathcal{Z}_1=-4\frac{d^3}{dx^3}+ 6U_1\frac{d}{dx}+3U_{1x}$ are
the Lax operators corresponding to  the first two evolutionary
equations from the KdV hierarchy, $u_t-u_x=0$ and
$u_t-6uu_x+u_{xxx}=0$. Relation $[\Z_1,H_1]=0$ reduces here to the
Novikov equation of the form
$-\frac{1}{4}(U_{1xx}-3U_1^2-4\kappa_1^2U_1)_x=0$, which is
satisfied due to the equality
\begin{equation}\label{U1iden}
    U_{1xx}-3U_1^2-4\kappa_1^2U_1=0\,,
\end{equation}
valid for the one-soliton potential (\ref{V1sol}).

By virtue of relations (\ref{JJH})  and (\ref{ZZH}), the composed
system, described by the matrix $2\times 2$ Hamiltonian
$\mathcal{H}_n=\text{diag}\,(H_n\,,\tilde{H}_n)$, possesses six
nontrivial self-adjoint integrals
\begin{equation}\label{SQP}
\mathcal{S}_{n,1}=
\left(
\begin{array}{cc}
 0 & \X_n   \\
 \X_n^\dagger &   0
\end{array}
\right),
\quad
\mathcal{Q}_{n,1}=
\left(
\begin{array}{cc}
 0 & \Y_n   \\
 \Y_n^\dagger &   0
\end{array}
\right),\quad
\mathcal{P}_{n,1}=
-i\left(
\begin{array}{cc}
 \Z_n & 0   \\
 0 &   \tilde{\Z}_n
\end{array}
\right),
\end{equation}
and $\mathcal{S}_{n,2}=i \sigma_3\mathcal{S}_{n,1}$,
$\mathcal{Q}_{n,2}=i \sigma_3\mathcal{Q}_{n,1}$,
$\mathcal{P}_{n,2}= \sigma_3\mathcal{P}_{n,1}$. A choice of the
diagonal Pauli sigma matrix $\sigma_3$ as the $\Z_2$-grading
operator identifies  integrals $\mathcal{S}_{n,a}$ and
$\mathcal{Q}_{n,a}$, $a=1,2$,  as the fermion operators,
$\{\sigma_3,\mathcal{S}_{n,a}\}=
\{\sigma_3,\mathcal{Q}_{n,a}\}=0$, while  $\mathcal{P}_{n,a}$ are
identified as the boson ones, $[\sigma_3,\mathcal{P}_{n,a}]=0$.
Together with $\mathcal{H}_n$ they generate a nonlinear
superalgebra, in which the Hamiltonian $\mathcal{H}_n$ plays a
role of the multiplicative central charge. The superalgebraic
structure given by the anti-commutation  relations of these
integrals, whose explicit form can be found in \cite{PAM}, is
insensitive to translation parameters $\tau_j$ and
$\tilde{\tau}_j$. Here we only write down the explicit form
of the commutation relations of the bosonic integrals with the
supercharges,
\begin{equation}\label{PSQn}
    [\mathcal{P}_{n,1},\mathcal{S}_{n,a}]=
    i\mathcal{H}_{n}\P^-_n(\mathcal{H}_n,\kappa\,,
    \tilde{\kappa})
     \mathcal{Q}_{n,a},\qquad
    [\mathcal{P}_{n,1},\mathcal{Q}_{n,a}]=
    -i\P^-_n(\mathcal{H}_n,\kappa,
    \tilde{\kappa})\mathcal{S}_{n,a}\,,
\end{equation}
and the commutators for $\mathcal{P}_{n,2}$
have a similar form but with
$\P^-_n(\mathcal{H}_n,\kappa,\tilde{\kappa})
$ changed
for $\P^+_n(\mathcal{H}_n,\kappa,
    \tilde{\kappa})$,
    where
$\P^\pm_n(\mathcal{H}_n,\kappa,\tilde{\kappa})\equiv
\P_n(\mathcal{H}_n,\kappa)\pm
\P_n(\mathcal{H}_n,\tilde{\kappa})$,
and
\begin{equation}\label{PndefProd}
\P_n(\mathcal{H}_n,\kappa)=
\prod_{j=1}^n(\mathcal{H}_n+\kappa_j^2\mathds{1})\,,
\end{equation}
with $\mathds{1}$ to be a unit $2\times 2$ matrix. {}From
definition of $\P^\pm_n$ it follows that while the $\P^+_n$ is
always a polynomial of order $n$ in the matrix Hamiltonian
$\mathcal{H}_n$, the $\P^-_n$ in a generic case is a polynomial of
order $(n-1)$ in $\mathcal{H}_n$. Moreover, in a completely
isospectral case given by the conditions
$\kappa_j=\tilde{\kappa}_j$, $j=1,\ldots, n$, $\P^-_n$ reduces to
the zero operator. This means that in such a completely
isospectral case the bosonic integral $\mathcal{P}_{n,1}$
transmutes into the central charge of the nonlinear superalgebra.
In the next section we show that the  family of the  systems
$\mathcal{H}_n$ composed from completely isospectral pairs $H_n$
and $\tilde{H}_n$ with pairwise coinciding bound states energies
$E_j=\tilde{E}_j=-\kappa^2_j$, $j=1,\ldots,n$, contains a special
subset of Schr\"odinger supersymmetric systems in which the
supercharges $\mathcal{S}_{n,a}$, $a=1,2$, of differential order
$(2n+1)$ are reduced to the two supercharges to be matrix
differential operators of the first order.

\section{Special  family of  isospectral $n$-soliton systems
and their exotic supersymmetry}

The described intertwining operators and, as a consequence, fermionic
integrals for the extended system $\mathcal{H}_n$  are irreducible as
soon as all the discrete energy levels of the subsystem $H_n$ are
 different from those for the subsystem $\tilde{H}_n$, i.e. when
 $\kappa_j\neq \tilde{\kappa}_{j'}$ for any  values of $j$ and  $j'$,
 $j,j'=1,\ldots,n$.
As it was shown in \cite{PAM}, when  any $r$,  $0<r\leq n$,  discrete
energy levels of one subsystem coincide with any $r$  discrete
energy levels of another subsystem, one or both of the
intertwining operators (\ref{YXinter}) are reducible  in such a
way that the total order of the two basic intertwining generators
reduces to $4n-2r+1$. The superalgebraic structure acquires then a
dependence on the corresponding $r$ relative translation
parameters. As we have just seen,
the case of a complete pairwise coincidence of the
discrete energy levels, $\kappa_j=\tilde{\kappa}_j$, $j=1,\ldots,
n$, is detected by transformation of the bosonic integral
$\mathcal{P}_{n,1}$ into the
 central charge of the
 superalgebra. It was also made an observation in \cite{PAM}
 that within such a class of the
 systems,   there is  a special, infinite
 family $\mathcal{H}_n$, $n=1,2,\ldots$,
  such that the corresponding
  completely   isospectral reflectionless partners
  $H_n$ and $\tilde{H}_n$
 are intertwined  by the first order differential
 operators  ${X}_n$ and $X_n^\dagger$.
 The first order intertwiners $X_n$ and $X_n^\dagger$
   replace
  the reducible operators $\X_n$ and $\X_n^\dagger$ of the odd
  order $2n+1$, while
 the intertwining generators $\Y_n$  and $\Y_n^\dagger$
 of the even order $2n$
 remain to be the same  as  in (\ref{YXinter}).
 More precisely, in \cite{PlyNi,PAM} it was found that the
 reflectionless systems $H_1=H_1(\kappa_1,\tau_1)$  and
 $\tilde{\mathcal{H}}_1=H_1(\kappa_1,\tilde{\tau}_1)$ can be related
  by the first order intertwining operators $X_1$ and $X_1^\dagger$,
\begin{equation}\label{X1}
    X_1=\frac{d}{dx}+\Omega_1-\tilde{\Omega}_1+\mathcal{C}_1\,,
\end{equation}
so that
\begin{equation}\label{X11H}
    X_1X_1^\dagger=H_1+\mathcal{C}_1^2,\qquad
    X_1^\dagger X_1=\tilde{H}_1+\mathcal{C}_1^2\,,
\end{equation}
where $\tilde{\Omega}_1=\Omega(\kappa_1,\tilde{\tau}_1)$ and
\begin{equation}\label{C1}
    \mathcal{C}_1=\kappa_1\coth
    \kappa_1(\tau_1-\tilde{\tau}_1)\,.
\end{equation}
Similarly, for the next case of $n=2$,  completely isospectral
Hamiltonians $H_2$ and $\tilde{H}_2$ satisfy the relations
$X_2X_2^\dagger=H_1+\mathcal{C}_2^2$ and  $X_2^\dagger
X_2=\tilde{H}_2+\mathcal{C}_2^2$ if
$\varphi_1\equiv\tau_1-\tilde{\tau}_1$ is fixed in terms of
$\varphi_2\equiv\tau_2-\tilde{\tau}_2\neq 0$ by a condition
$\mathcal{C}_1=\mathcal{C}_2$, where  $\mathcal{C}_2$ and $X_2$
are given by relations of the form (\ref{C1})  and (\ref{X1}) with
the  index $1$ changed for $2$. Based on these
two special cases with $n=1$ and $n=2$,
it was conjectured  in \cite{PAM}  that
such a picture with the first order intertwining generators
can be
generalized  for the case of arbitrary $n$.

We will show now that any two completely isospectral
reflectionless Hamiltonians $H_n$ and $\tilde{H}_n$ with
translation parameters constrained by a condition
\begin{equation}\label{CCC}
    \mathcal{C}_1=\mathcal{C}_2=\ldots=
    \mathcal{C}_n=\mathcal{C}\,,
\end{equation}
are indeed related by
the first order operators  $X_n$ and $X_n^\dagger$,
\begin{equation}\label{Xn}
    X_n=\frac{d}{dx}+\Omega_n-\tilde{\Omega}_n+\mathcal{C}\,,
\end{equation}
\begin{equation}\label{XnnH}
    X_nX_n^\dagger=H_n+\mathcal{C}^2,\qquad
    X_n^\dagger X_n=\tilde{H}_n+\mathcal{C}^2\,,
\end{equation}
\begin{equation}\label{XnHinter}
    X_n^\dagger H_n=\tilde{H}_nX_n^\dagger\,,
    \quad
    X_n\tilde{H}_n=H_nX_n\,,
\end{equation}
where $\mathcal{C}_j=\kappa_j
\coth\kappa_j(\tau_j-\tilde{\tau}_j)$, and $\mathcal{C}$ is a real
parameter restricted by inequality $
\mathcal{C}^2>\kappa_n^2$.

To prove the validity of the statement, we first rewrite
Eq. (\ref{C1}) in the form
\begin{equation}\label{Om1C}
    \mathcal{C}_1=\frac{\Omega_1\tilde{\Omega}_1-\kappa_1^2}{\Omega_1-
    \tilde{\Omega}_1}\,
\end{equation}
by using the elementary identity $\coth (\alpha-\beta)=(
1-\tanh\alpha\tanh\beta)/(\tanh\alpha-\tanh\beta)$. The chain of
constraints (\ref{CCC}) can be presented equivalently as
\begin{equation}
    \mathcal{C}_j=\frac{\Omega_{1}(j)\tilde{\Omega}_{1}(j)
    -\kappa_j^2}{\Omega_{1}(j)-
        \tilde{\Omega}_{1}(j)}=\mathcal{C}\,,\qquad
    j=1,\ldots,n\,.
\end{equation}
 Relations
(\ref{Xn}) and (\ref{XnnH}) imply  two equalities
\begin{eqnarray}\label{XO}
    U_n+\mathcal{C}^2&=&(\Omega_n-\tilde{\Omega}_n+\mathcal{C})^2
    +(\Omega_n-\tilde{\Omega}_n)_x\,,\\
    \tilde{U}_n+\mathcal{C}^2&=&(\Omega_n-\tilde{\Omega}_n+\mathcal{C})^2
    -(\Omega_n-\tilde{\Omega}_n)_x\,.\label{XOt}
\end{eqnarray}
To prove (\ref{XnnH}) under condition (\ref{CCC}) we have to
demonstrate the validity of relations  (\ref{XO}) and (\ref{XOt}).
A difference of these two relations gives
$U_n-\tilde{U}_n=2(\Omega_n-\tilde{\Omega}_n)_x$, that is true
because of (\ref{omV}). Denoting  $\mathcal{U}_n\equiv
U_n+\tilde{U}_n- 2 [(\Omega_n-\tilde{\Omega}_n
+\mathcal{C})^2-\mathcal{C}^2]$, we have to show that
$\mathcal{U}_n=0$. {}For   $n=1$, the equality $\mathcal{U}_1=0$
is checked directly by using (\ref{V1sol}) and (\ref{C1}). Then it
is sufficient to prove that $\mathcal{U}_n+ \mathcal{U}_{n-1}=0$
for any $n>1$. {}From Eq.  (\ref{UnUn-1}) we have
$U_n+U_{n-1}-2[(\Omega_n-\Omega_{n-1})^2-\kappa_n^2]=0$. Let us
add this last equality  to its analog obtained by changing
$\tau_j\rightarrow\tilde{\tau}_j$. Subtracting the obtained left
hand side expression (equal to zero) from $\mathcal{U}_n+
\mathcal{U}_{n-1}$,  we arrive finally at the equality
\begin{equation}\label{Gamman}
    \Gamma_n\equiv-\kappa_n^2+(\Omega_n-\tilde{\Omega}_{n-1})
    (\tilde{\Omega}_n-\Omega_{n-1})-\mathcal{C}(\Omega_n+\Omega_{n-1}-
    \tilde{\Omega}_n-\tilde{\Omega}_{n-1})=0\,,
\end{equation}
which has to be proved. Remembering that
$\Omega_0=\tilde{\Omega}_0=0$, the validity
of $\Gamma_1=0$ follows
from (\ref{Om1C}). Assume that the relation $\Gamma_n=0$ is valid
for arbitrary  value of $n>1$ with any admissible  set of
parameters satisfying the constraint (\ref{CCC}). Some algebraic
manipulations with employing recursive relation (\ref{Omegarec})
with $n\rightarrow n+1$ gives rise to the equality
 \begin{equation}
    \Gamma_{n+1}=\Gamma_n +\frac{\kappa_{n+1}^2-\kappa_{n}^2}{
    (\Omega_n-\Omega_n^\sharp)(
    \tilde{\Omega}_n-\tilde{\Omega}_n^\sharp)}
    \left(\Gamma_n-\Gamma_n^\sharp\right),
\end{equation}
where $\Gamma_n^\sharp$ is obtained from $\Gamma_n$ by changing in
(\ref{Gamman}) $\Omega_n$ and $\tilde{\Omega}_n$ for
$\Omega_n^\sharp$ and $\tilde{\Omega}_n^\sharp$, and $\kappa_n$
for $\kappa_{n+1}$,
 and we have taken into
account here constraint (\ref{CCC}) extended for the case
$n\rightarrow n+1$. The equality $\Gamma_{n+1}=0$ follows then
from $\Gamma_n=0$, that proves the validity of relations
(\ref{XnnH}) for completely isospectral pairs of reflectionless
$n$-soliton Hamiltonians with translation parameters constrained
by the condition (\ref{CCC}).

\begin{figure}[h!]\begin{center}
    \includegraphics[scale=0.6]{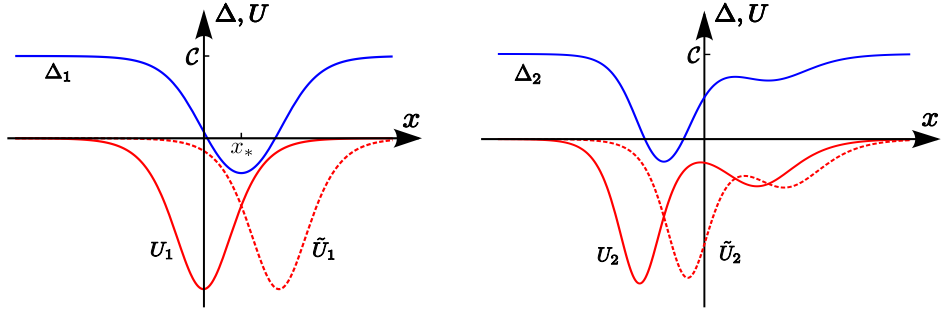}
    \caption{ \emph{On the left}: One-soliton isospectral
     potentials of the P\"oschl-Teller form (\ref{V1sol}),
    and corresponding superpotential (kink-anti-kink
    background)  $\Delta_1$.
    Relative phase $\tau_1-\tilde{\tau}_1$ coincides
    here with the distance between minima of potentials
    $\tilde{U}_1$ and $U_1$, and defines the asymptotic
    value $\mathcal{C}$  of the superpotential
    via the relation (\ref{C1}).
    The  $\Delta_1(x)$
    has a mirror-symmetric
    form,
    and $n=1$ is a unique case
    for which $\tilde{U}_n$ is a
    simple translation of $U_n$.
    \emph{On the right}: Isospectral two-soliton potentials
    of the
    general form (\ref{U2}), with
    relative soliton phases subjected
    to the condition (\ref{CCC}), or,
    equivalently, with the translation
    parameters shifts  related by Eq. (\ref{phij}).
    The values of potentials coincide in three
    different points, where
    the  corresponding  superpotential
    $\Delta_2$
    has three extrema,
    two of which correspond to two local minima.
    Note that   $\Delta_n$ with $n>1$  not
    obligatorily has  $n$ minima.  For instance,
    the potential $U_2$
    with  $\kappa_2=2\kappa_1$ and $\tau_2=\tau_1$
    will have a symmetric form, similar to the form
    (\ref{V1sol}) of the one-soliton
    P\"oschl-Teller potential $U_1$  with the amplitude
    coefficient $-2\kappa_1^2$
     changed for $-6\kappa_1^2$.
    In this case the isospectral potential
    $\tilde{U}_2$ with translation parameters given
     by Eq. (\ref{phij})
     will have a form different from that of $U_2$
     and will coincide with it only in one point, and
    corresponding superpotential
    (reflectionless Dirac potential, see below)
    $\Delta_2$ will have
    only one minimum.}
    \label{fig1}
\end{center}
\end{figure}

Condition (\ref{CCC}) allows us to fix the shifts
$\varphi_j=\tau_{j}-\tilde{\tau}_{j}$, $j=1,\ldots, n,$ in terms
of the free parameter $\mathcal{C}$, $\mathcal{C}^2> \kappa_n^2$,
and $\kappa_j$,
\begin{equation}\label{phij}
    \varphi_j(\kappa_j,\mathcal{C})=\frac{1}{2\kappa_j}\ln
    \frac{\mathcal{C}+\kappa_j}{\mathcal{C}-\kappa_j}=
    \frac{1}{\kappa_j}\text{arctanh}(\kappa_j/\mathcal{C})\,.
\end{equation}
A reflectionless system $\mathcal{H}_n$ from the special infinite
family characterized by the properties (\ref{CCC}), (\ref{Xn}),
(\ref{XnnH}) and  (\ref{XnHinter}) is given therefore by $2n+1$
parameters. Denoting
\begin{equation}\label{defDeltan}
    \Delta_n(x)=\Delta_n(x;\kappa_j,\tau_j,\mathcal{C})\equiv
    \Omega_n\left(\kappa_j(x+\tau_j)\right)-
    \Omega_n\left(\kappa_j(x+\tau_j-\varphi_j(\kappa_j,\mathcal{C}))
    \right)
    +\mathcal{C}\,,
\end{equation}
we present the first order intertwining operator
 in the  form
\begin{equation}\label{XnDelta}
    X_n(x;\kappa_j,\tau_j,\mathcal{C})=\frac{d}{dx}+
    \Delta_n(x;\kappa_j,\tau_j,
    \mathcal{C})\,,
\end{equation}
that can be compared with
the structure of the first order
differential operator
$A_n=\frac{d}{dx}+\Omega_n-\Omega_{n-1}=\frac{d}{dx}+
\mathcal{W}_n$. In correspondence with (\ref{XO}), we have
\begin{equation}\label{Delkapt}
    \Delta_n^2+
    \Delta_{nx}=U_n+\mathcal{C}^2\,,\qquad
    \Delta_n^2-
    \Delta_{nx}=\tilde{U}_n+\mathcal{C}^2\,,
\end{equation}
cf. (\ref{VVW}). It is worth to note here that the change
\begin{equation}\label{C-Ctt}
    \mathcal{C}\rightarrow -\mathcal{C},
    \qquad
     \tau_j\rightarrow \tau_j
    -\varphi_j(\kappa_j,\mathcal{C})
    =\tilde{\tau}_j
\end{equation}
induces the changes $\Delta_n(x)\rightarrow -\Delta_n(x)$,
$U_n(x)\leftrightarrow\tilde{U}_n(x)$. Note also  that  the points
$x_*$  where the values of potentials $U_n$ and $\tilde{U}_n$
coincide, $U_n=\tilde{U}_n$, correspond in general to local
extrema of $\Delta_n$, see Figure \ref{fig1} illustrating the
cases  $n=1$ and $n=2$ with $\mathcal{C}>0$.

The first order operators $X_n$, $A_n$ and $\tilde{A}_n$ satisfy
the intertwining  relation
\begin{equation}\label{AXn}
    A_nX_{n-1}=X_n\tilde{A}_n\,.
\end{equation}
To show this, we note that
 (\ref{AXn}) is equivalent to the equality
\begin{equation}\label{COnx}
    (\mathcal{C}+\Omega_{n-1}-\tilde{\Omega}_n)
    (\Omega_n-\tilde{\Omega}_n-\Omega_{n-1}+\tilde{\Omega}_{n-1})
    =
    (\tilde{\Omega}_{n}-\Omega_{n-1})_x\,.
\end{equation}
By virtue of relations $U_n=2\Omega_{nx}$ and  (\ref{XOt}), and
the second equality from (\ref{VVW}), we have
\begin{equation}
    (\tilde{\Omega}_{n}-\Omega_{n-1})_x
    =(\Omega_{n-1}-\tilde{\Omega}_n)(2\Omega_n-\tilde{\Omega}_n
    -\Omega_{n-1})+2\mathcal{C}(\Omega_n-\tilde{\Omega}_n)+\kappa_n^2\,.
\end{equation}
As a consequence,  (\ref{COnx}) is reduced to the equality
$\Gamma_n=0$, see Eq. (\ref{Gamman}), that proves
validity of   relation
(\ref{AXn}). Having in mind the definition (\ref{XnnH}),
intertwining relation $X_n\tilde{H}_n=H_nX_n$, and that
$H_n=H_n(\kappa_j,\tau_j)$,
$\tilde{H}_n=H_n(\kappa_j,\tilde{\tau}_j)$,
 it is convenient to write  $X_n=X_n(\tau_j,\tilde{\tau_j})$.
Then $X_n^\dagger(\tau_j,\tilde{\tau}_j)=
 -X_n(\tilde{\tau}_j,\tau_j)$, and
a conjugation of (\ref{AXn}) after the change
$\tau_j\leftrightarrow \tilde{\tau}_j$ gives us also the
intertwining relation
\begin{equation}\label{AXn+}
    X_{n-1}\tilde{A}_n^\dagger=A_n^\dagger X_n\,.
\end{equation}

Using  (\ref{AXn}) and  (\ref{AXn+}), we find that the
intertwining operator  $\X_n$  of order $2n+1$ defined in
(\ref{YXinter}), in the present case of the special isospectral
pairs of the Hamiltonians reduces as
\begin{equation}
    \X_n=-\mathcal{C}\Y_n+\prod_{j=1}^n(H_n+\kappa_j^2)\cdot X_n\,.
\end{equation}
Equivalently, this can be presented in the form
\begin{equation}\label{X0Xn}
    \A_nX_0\tilde{\A}^\dagger_n=
    \prod_{j=1}^n(H_n+\kappa_j^2)\cdot X_n\,,
    \qquad X_0=\frac{d}{dx}+\mathcal{C}\,.
\end{equation}
Eq. (\ref{X0Xn}) shows that  the intertwining operator  $X_n$
 is a Darboux-dressed form of the operator $X_0$. The operator
 $X_0$  intertwines
  the Hamiltonian $H_0$ of the free  particle  with itself,
  $[X_0,H_0]=0$.

Because of  the reducible character of the operator $\X_n$,
 integrals $\mathcal{S}_{n,a}$  of the system
  $\mathcal{H}_n={\rm diag}\,(H_n,\tilde{H}_n)$
  from the special family we consider are  also reducible,
$\mathcal{S}_{n,a}=-\mathcal{C}\mathcal{Q}_{n,a}
+\prod_{j=1}^n(\mathcal{H}_n +\kappa_j^2)\breve{\mathcal{S}}_{n,a}$,
where
$\breve{\mathcal{S}}_{n,a}$
 have a structure like in (\ref{SQP}) but
with differential operators $\X_n$ and $\X_n^\dagger$  of the
order $2n+1$ changed for the first order operators $X_n$ and
$X_n^\dagger$. The nontrivial integrals
$\breve{\mathcal{S}}_{n,a}$, $\mathcal{Q}_{n,a}$ and
$\mathcal{P}_{n,a}$
generate together with the Hamiltonian $\mathcal{H}_n$ a nonlinear
superalgebra with the following nontrivial (anti)-commutation
relations,
\begin{equation}\label{susyex1}
    \{\breve{\mathcal{S}}_{a},
    \breve{\mathcal{S}}_{b}\}=2\delta_{ab}(\mathcal{H}_n+
    \mathcal{C}^2),\quad
    \{\mathcal{Q}_{a},\mathcal{Q}_{b}\}=
    2\delta_{ab}\P_n^2\,,\quad
    \{\breve{\mathcal{S}}_{a},\mathcal{Q}_{b}\}=
    2\delta_{ab}\mathcal{C}\P_n
    +2\epsilon_{ab}\mathcal{P}_{1}\,,
\end{equation}
\begin{equation}\label{susyex2}
    [\mathcal{P}_{2},\breve{\mathcal{S}}_{a}]=
    2i\left((\mathcal{H}_n+\mathcal{C}^2)
    \mathcal{Q}_{a}
    -\mathcal{C}\P_n\breve{\mathcal{S}}_{a}\right)\,,\quad
    [\mathcal{P}_{2},\mathcal{Q}_{a}]=
    2i\P_n\left(\mathcal{C}\mathcal{Q}_{a}
    -\P_n\breve{\mathcal{S}}_{a}\right)\,,
\end{equation}
where $\P_n=\P_n(\mathcal{H}_n,\kappa)$ is the operator
defined in Eq. (\ref{PndefProd}), and to simplify
expressions, we omitted index $n$ in notation of the integrals
$\breve{\mathcal{S}}_{n,a}$, $\mathcal{Q}_{n,a}$ and
$\mathcal{P}_{n,a}$. Integral $\mathcal{P}_{n,1}$ commutes with
all other integrals, and
 plays a role of the central charge of the
superalgebra.

As follows from (\ref{susyex1}),  (\ref{PndefProd}), and relation
$\mathcal{C}^2>\kappa_j^2$, the first order supercharges
$\breve{\mathcal{S}}_{a}$ are the positive definite operators, and
the part of supersymmetry associated with them is spontaneously
broken. According to the first relation from (\ref{susyex1}), the
kernels of these two supercharges are formed by non-physical
eigenstates of  $\mathcal{H}_n$. On the other hand, each of the
two supercharges $\mathcal{Q}_{a}$ detects all the doubly
degenerate discrete eigenvalues of $\mathcal{H}_n$ by annihilating
all the $2n$ bound states of the matrix Hamiltonian operator. The
central supercharge $\mathcal{P}_{n,1}$, generated via the
anticommutation of supercharges $\breve{\mathcal{S}}_{a}$ and
$\mathcal{Q}_{b}$ with $a\neq b$, annihilates not only all the
bound states, but also detects two zero energy  states at the edge
of the continuum part of the spectrum of $\mathcal{H}_n$ by
annihilating them. The rest of the continuous part of the spectrum
of $\mathcal{H}_n$ with $E>0$ is the fourth-fold degenerate. The
second nontrivial bosonic integral, $\mathcal{P}_{n,2}$, not
appearing in the  anticommutation relations of the supercharges,
plays a role of the operator acting on the pairs of supercharges
$(\breve{\mathcal{S}}_{a},\,\mathcal{Q}_{a})$ with $a=1$ and $a=2$
as a rotation type operator. Note that from (\ref{susyex2}) it
follows, particularly, that $[\mathcal{P}_{2},\mathcal{C}Q_{a}
    -\P_n\breve{\mathcal{S}}_{a}]=
    -2i\P_n\mathcal{H}_nQ_{a}$.

\section{Dirac reflectionless systems and the mKdV solitons}

Let us look at the obtained results  from a completely
 different, though related,  perspective.
 Take one of the two  integrals
 $\breve{\mathcal{S}}_{n,a}$,
 say $\breve{\mathcal{S}}_{n,1}$, and identify it as a
 Dirac type Hamiltonian,
 \begin{equation}\label{HDir}
\mathcal{H}_n^D=
\left(
\begin{array}{cc}
 0 & X_n   \\
 X_n^\dagger &   0
\end{array}
\right) =\left(
\begin{array}{cc}
 0 & \partial_x+\Delta_n   \\
 -\partial_x+\Delta_n &   0
\end{array}
\right).
\end{equation}
According to Eq.  (\ref{X0Xn}), in the case $n=0$ operator
(\ref{HDir})  describes a free Dirac particle of the mass
$\vert\mathcal{C}\vert$, $\mathcal{H}_0^D=
-i\frac{d}{dx}\sigma_2+\mathcal{C}\sigma_1$, while
$\mathcal{H}_n^D$ with $n\geq 1$ is a Darboux-dressed form of
$\mathcal{H}_0^D$, $\mathcal{A}_n \mathcal{H}_0^D
\mathcal{A}_n^\dagger=\mathcal{H}_n^D \prod_{j=1}^n
\left((\mathcal{H}^D_n)^2 +(\kappa_j^2-\mathcal{C}^2)
\mathds{1}\right)$, where $\mathcal{A}_n=\text{diag}\,
(\A_n,\tilde{\A}_n)$, see Eq. (\ref{X0Xn}). In last section it
will be indicated that the first order matrix reflectionless
operator $\mathcal{H}_n^D$ can also be considered as the BdG
Hamiltonian in
 Andreev approximation \cite{Stone}.
Then function $\Delta_n(x)$ appearing in its structure has, in
dependence on a physical context, a meaning of a
gap function, a
condensate, an order parameter, or just
a position-dependent mass.
Note that relations (\ref{XnDelta}),
(\ref{phij}), (\ref{Omegarec}), (\ref{V1sol})
 and (\ref{Om123})  allow us
to construct $\Delta_n(x)$ recursively
for any $n$.

The Dirac reflectionless system  (\ref{HDir})
has a nontrivial matrix integral $\mathcal{P}_{n,1}$
 given by Eqs. (\ref{SQP}), (\ref{defZ}),
 which is a dressed form
of the linear momentum integral $-i\frac{d}{dx}\mathds{1}$ of the
free Dirac particle $\mathcal{H}_0^D$, $\mathcal{P}_{n,1}=
\mathcal{A}_n (-i\frac{d}{dx}\mathds{1}) \mathcal{A}_n^\dagger$.
The relation of commutativity
$[\mathcal{H}^D_n,\mathcal{P}_{n,1}]=0$, following immediately
from the Darboux-dressed nature of the matrix operators
$\mathcal{H}_n^D$ and $\mathcal{P}_{n,1}$ is equivalent to the
intertwining relation
\begin{equation}\label{ZnXn}
     \Z_nX_n=X_n\tilde{\Z}_n\,,
\end{equation}
and to the conjugate relation,
$X_n^\dagger\Z_n=\tilde{\Z}_nX_n^\dagger$, which follows from
(\ref{ZnXn}) under the change
$\tau_j\leftrightarrow\tilde{\tau}_j$. \vskip0.1cm

Consider as an example in more detail the simplest nontrivial case
$n=1$ \cite{DHN}. We have
\begin{equation}\label{defD1}
    \Delta=\kappa\left(-\tanh\kappa(x+\tau)+\tanh\kappa
    (x+\tilde{\tau})\right)+\mathcal{C},\qquad
     \mathcal{C}= \kappa\,{\rm
    coth}\,\kappa(\tau-\tilde{\tau})\,,
\end{equation}
and so, the sign of $\mathcal{C}$ coincides with the sign of
$(\tau-\tilde{\tau})$.
To simplify notations, we omitted here index $1$ in $\Delta$,
$\kappa$ and $\tau$. This gap function satisfies an ordinary
nonlinear differential equation
\begin{equation}\label{Dx1id}
    \Delta_x^2=(\Delta-\mathcal{C})^2(\Delta-\delta_+)
    (\Delta-\delta_-)\,,\quad
    {\rm where}\quad
    \delta_\pm=-\mathcal{C}\pm 2\sqrt{\mathcal{C}^2-\kappa^2}\,.
\end{equation}
 {}From (\ref{defD1})  it follows that  $\Delta(x)$
is even function with
respect to the point $x=x_*\equiv-\frac{1}{2}(\tau+\tilde{\tau})$,
where it takes a minimum (or maximum) value $\delta_+$ (or
$\delta_-$) for $\mathcal{C}>0$ ($\mathcal{C}<0$). Its form for
the case $\mathcal{C}>0$ is shown on Figure \ref{fig1}.

As a consequence
of (\ref{Dx1id}), $\Delta(x)$ satisfies also equations
\begin{equation}
    \Delta_{xx}=2\Delta^3+2\Delta(2\kappa^2-3\mathcal{C}^2)+4
    \mathcal{C}(\mathcal{C}^2-\kappa^2)\,,
\end{equation}
\begin{equation}\label{D3x}
    \Delta_{xxx}=6\Delta^2\Delta_x+2\Delta_x(2\kappa^2-3\mathcal{C}^2)\,.
\end{equation}

With taking into account relation (\ref{U1iden}), we find that for
$n=1$  the intertwining relation (\ref{ZnXn}) is equivalent, as a
condition of equality to zero of the coefficients at $d^2/dx^2$,
$d/dx$ and $1$,
 to the
three equations: $\,U-\tilde{U}=2\Delta_x$,
$2(U-\tilde{U})\Delta+(U_x-3\tilde{U}_x)= 4\Delta_{xx}$ and
$(6U+4\kappa^2)\Delta_x+
3(U_x-\tilde{U}_x)\Delta-3\tilde{U}_{xx}=4\Delta_{xxx}$.  The
first two of these equations are satisfied by virtue of
(\ref{Delkapt}). The third equation is then satisfied by taking
into account (\ref{U1iden}) and relation of the same form for
$\tilde{U}$.

Let us present equality  (\ref{D3x}) satisfied by the function
$\Delta=\Delta(x;\tau,\mathcal{C})$ in the form
$6\Delta^2\Delta_x-\Delta_{xxx}=
(6\mathcal{C}^2-4\kappa^2)\Delta_x$.
Assume now that $\Delta$ depends additionally
on an evolution parameter $t$
in such a way that $\Delta(x,t=0)=\Delta(x;\tau,\mathcal{C})$,
and fix such a dependence in the form
\begin{equation}\label{Delta1xt}
    \Delta(x,t)\equiv\Delta(x+(6\mathcal{C}^2-4\kappa^2)t;
    \tau,\mathcal{C})\,.
\end{equation}
Then $\Delta_t=\Delta_x(6\mathcal{C}^2-4\kappa^2)$, and function
$\Delta(x,t)$ will satisfy the mKdV equation
$\Delta_t-6\Delta^2\Delta_x+
\Delta_{xxx}=0$. Equation
(\ref{D3x}) in this case
will be  a stationary equation of the mKdV hierarchy.

The described  observation can be generalized for  the case of
arbitrary $n$. For this we first note that if
$U_n(x;\kappa_j,\tau_j)$ is a general $2n$-parametric $n$-soliton
potential constructed in accordance with the inverse scattering
method for $t=0$, the dependence on $t$ in correspondence with the
KdV equation is obtained by the the substitution
$\tau_j\rightarrow -4\kappa_j^2t+\tau_j^0$, where $\tau_j^0$,
$j=1,\ldots,n$, are constant parameters. The KdV equation
possesses Galilean symmetry: if $u(x,t)$ is a solution of the KdV
equation, then $U(x,t)=u(x+6ct,t)+c$ is also solution for any
value of a constant $c$. Let us make a shift $x\rightarrow x+6ct$
in both sides of two relations in (\ref{Delkapt}), and rewrite the
obtained right hand sides in equivalent forms
$(U_n(x+6ct)+c)-c+\mathcal{C}^2$ and
$(\tilde{U}_n(x+6ct)+c)-c+\mathcal{C}^2$. Put now
$c=\mathcal{C}^2$ and denote
$U_n(x+6\mathcal{C}^2t)+\mathcal{C}^2=u^+(x,t)$,
$\tilde{U}_n(x+6\mathcal{C}^2t)+\mathcal{C}^2=u^-(x,t)$,
$\Delta_n(x+6\mathcal{C}^2t)=v(x,t)$. Exploiting  then a  relation
between the KdV and the mKdV equations, which is described in
 Appendix B, we conclude that
the function
\begin{equation}\label{Deltanxt}
    \Delta_n(x,t)=\Omega_n(\xi_j)-
    \Omega_n(\tilde{\xi}_j)+\mathcal{C}
    \end{equation}
where  $ \xi_j=\kappa_j(x+(6\mathcal{C}^2-4
\kappa_i^2)t +\tau_j^0)$,
    $ \tilde{\xi}_j=\xi_j-\kappa_j\varphi_j(\kappa_j,\mathcal{C})$,
    $j=1,\ldots,n$,
is the $n$-soliton solution of the mKdV equation
$v_t-6v^2v_x+v_{xxx}=0$.
 In particular case $n=1$, Eq.  (\ref{Deltanxt})
 corresponds to (\ref{Delta1xt}).

\section{Fermion system in a multi-kink-antikink background
as a Darboux-dressed free massive Dirac particle}

Here we show that the reflectionless Dirac system described by the
first order matrix Hamiltonian (\ref{HDir}),
or that is the same, a fermion system in a
multi-kink-anti-kink background, possesses its own
exotic supersymmetry
that is rooted in the peculiar supersymmetry
of the associated Schr\"odinger system studied in Section 3. It
can be understood as a dressed form of the supersymmetric
structure of the free massive Dirac particle.
This also will allow us to present
the trapped configurations (bound states) and scattering
states of our fermion system in an explicit analytic form.

Consider a free Dirac massive particle described by the
Hamiltonian
\begin{equation}\label{H0D}
\mathcal{H}^D_0=\left(
\begin{array}{cc}
  0& \partial_x+\mathcal{C}  \\
  -\partial_x+\mathcal{C} &   0
\end{array}
\right).
\end{equation}
 Its eigenfunctions
and corresponding eigenvalues are
\begin{equation}\label{freeeig}
    \Psi_{0,\pm}^k(x)=
\left(
\begin{array}{c}
 e^{ikx}\\
  \pm e^{ik(x+\varphi(k,\mathcal{C}))}
\end{array}
\right),\qquad
 \mathcal{E}_\pm(k)=\pm
    \sqrt{\mathcal{C}^2+k^2}\,.
\end{equation}
Here
\begin{equation}\label{varphik}
    \varphi(k,\mathcal{C})=\frac{1}{2ik}
    \ln\frac{\mathcal{C}-ik}{\mathcal{C}+ik}
\end{equation}
is  the function even in $k$,
 $\varphi(-k,\mathcal{C})=
 \varphi(k,\mathcal{C})$,
and odd in  $\mathcal{C}$, $\varphi(k,-\mathcal{C})=
 -\varphi(k,\mathcal{C})$, and  the quantity
 $e^{ik\varphi(k,\mathcal{C})}=
 \frac{\mathcal{C}-ik}{\sqrt{
\mathcal{C}^2+k^2}}$ is a pure phase, $\vert
e^{ik\varphi(k,\mathcal{C})}\vert=1$. The wave numbers $+k$ and
$-k$, $k>0$, correspond to the same, doubly degenerate energy
value. The plane wave states (\ref{freeeig}) with $k>0$ and $k<0$
are distinguished by the momentum integral
$-i\frac{d}{dx}\mathds{1}$. The eigenvalues
$\mathcal{E}_\pm(0)=\pm\vert \mathcal{C}\vert$ at the edges of the
upper and lower continuous bands are non-degenerate. The interval
$-\vert\mathcal{C}\vert<\mathcal{E}< \vert\mathcal{C}\vert$
corresponds to the energy gap in the spectrum of the free massive
Dirac particle.

Consider now the Dirac  reflectionless system
(\ref{HDir}). The Hamiltonian $\mathcal{H}^D_n$
anticommutes with
$\sigma_3$. Coherently with Eq. (\ref{freeeig}) corresponding to
the $n=0$ case, this implies  that if $\Psi_\mathcal{E}$ is an
eigenstate of $\mathcal{H}^D_n$, $\mathcal{H}^D_n\Psi_\mathcal{E}=
\mathcal{E}\Psi_\mathcal{E}$, then $\sigma_3\Psi_\mathcal{E}$ is
an eigenstate of eigenvalue $-\mathcal{E}$.

\begin{figure}[h!]\begin{center}
    \includegraphics[scale=0.6]{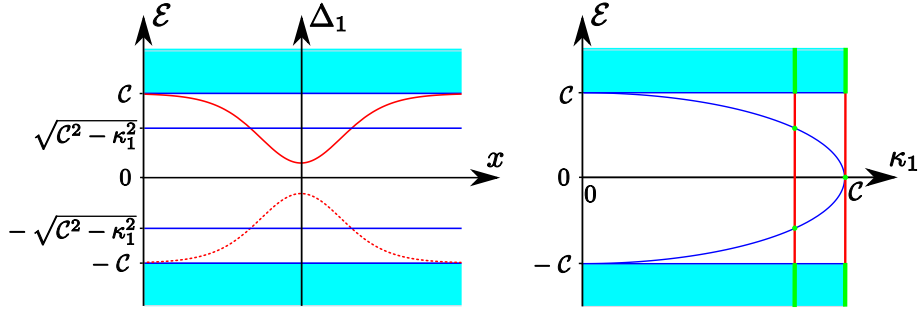}
    \caption{ \emph{On the left}:
    The form of reflectionless Dirac potential
    $\Delta_1(x)$ is shown by continuous curve
    for the case  $\mathcal{C}>0$,
    while dashed curve corresponds to
    isospectral kink-antikink background
     $-\Delta_1(x)$
    obtained by means of (\ref{C-Ctt}).
     Horizontal lines show two non-degenerate
     energy levels $\mathcal{E}=
     \pm\sqrt{\mathcal{C}^2
     -\kappa_1^2}$ of the bound states,
     and two non-degenerate energy levels
     $\mathcal{E}=
     \pm\mathcal{C}$
     at the edges of the doubly degenerate
     continuum part of the
     spectrum with $\mathcal{E}>
     \mathcal{C}$ and
     $\mathcal{E}<-
     \mathcal{C}$.
     \emph{On the right}:
     Dependence of  the spectrum for
    reflectionless Dirac system $\mathcal{H}^D_1$
    on the parameter $\kappa_1$.
    The curves correspond to the
    discrete energy levels
    $
    \pm\sqrt{\mathcal{C}^2-\kappa_1^2}$
    of the bound states.
    Two nonzero energy levels at
    $0<\kappa_1<\mathcal{C}$
    transform into one zero energy level in the limit case
     $\kappa_1=\mathcal{C}$.
     Kink-antikink background described by $\Delta_1(x)$
     transforms into an
     antitank background $\mathcal{W}_1=-\kappa_1
     \tanh\kappa_1(x+\tau_1)$ in the indicated limit
     (see also the discussion in last Section). In another limit,
     $\kappa_1\rightarrow 0$, $\mathcal{C}=const$, $\Delta_1(x)$
     transforms into the
     homogeneous background $\Delta_0=\mathcal{C}$.  }
    \label{fig2}
\end{center}
\end{figure}

\begin{figure}[h!]\begin{center}
    \includegraphics[scale=0.6]{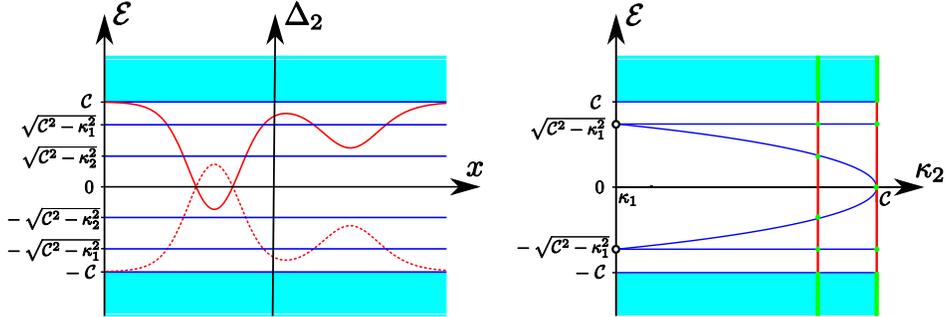}
    \caption{\emph{On the left}: The form of the potential
    and corresponding spectrum of the
    reflectionless Dirac system $\mathcal{H}^D_2$
    with the same notations as in Figure \ref{fig2}.
    \emph{On the right}:
    Spectrum of reflectionless
    Dirac system $\mathcal{H}^D_2$ in dependence of
    the parameter $\kappa_2$ varying in the interval
    $\kappa_1<\kappa_2<\mathcal{C}$.
    In the limit case $\kappa_2=\kappa_1$,
    inhomogeneous
    reflectionless Dirac potential $\Delta_2$ transforms into
    the homogenous background $\Delta_0=\mathcal{C}$,
    see Eqs. (\ref{Om123})
    and (\ref{defDeltan}), and the corresponding discrete energy
    levels $\pm\sqrt{\mathcal{C}^2-\kappa_1^2}$,
    shown by  blank
    circles,  disappear
    from the spectrum. In
    another limit $\kappa_2=\mathcal{C}$,
    $\mathcal{H}^D_2$  transforms
    into a reflectionless Dirac
    system with three non-degenerate energy levels,
    one of which has zero value. The square
    of $\mathcal{H}^D_2$  in this second limit
    gives a pair of almost isospectral
    reflectionless Schr\"odinger systems,
    one of which is described by a two-soliton potential
    supporting the bound state of zero energy,
    the potential of the other subsystem is one-soliton
    with nonzero energy of the bound state. }
    \label{fig3}
\end{center}
\end{figure}

 The eigenstates from the upper and lower continuums
in the spectrum of $\mathcal{H}^D_n$ are obtained by
Darboux-dressing of the plane wave states (\ref{freeeig}) of the
free particle,
 $\Psi_{n,\pm}^k(x)=
\mathcal{A}_n\Psi_{0,\pm}^k(x)$,
$\mathcal{H}^D_n\Psi_{n,\pm}^k(x)=\mathcal{E}_\pm(k)
\Psi_{n,\pm}^k(x)$, where $\mathcal{A}_n$ is the
 diagonal $2\times
2$ matrix, $\mathcal{A}_n=\text{diag}\,(\A_n,\tilde{\A}_n)$. The
bound states of $\mathcal{H}^D_n$ are constructed by
Darboux-dressing of the appropriate non-physical eigenstates  from
the energy gap of $\mathcal{H}^D_0$. First,  we note that function
(\ref{varphik}) for pure imaginary values $k=i\kappa_j$,
$\kappa_j>0$,  reduces to the relative soliton shifts
$\varphi_j(\kappa_j,\mathcal{C})$ given by Eq.  (\ref{phij}).
Taking linear combinations of the states of the form
(\ref{freeeig}) with $k=+i\kappa_j$ and $k=-i\kappa_j$,
$j=1,\ldots, n$, $0<\kappa_1<\ldots \kappa_{n-1}<\kappa_n <
\vert\mathcal{C}\vert$, we construct the formal, non-physical
eigenstates of $\mathcal{H}_0^D$ of eigenvalues
$\mathcal{E}_{0,\pm}(j)= \pm\sqrt{\mathcal{C}^2-\kappa_j^2}$,
\begin{equation}\label{eigenj0D}
    \Psi_{0,\pm}^j(x)=
\left(
\begin{array}{c}
 \cosh\kappa_j(x+\tau_j)\\
  \pm \cosh\kappa_j(x+\tilde{\tau}_j)
\end{array}
\right)\,,\qquad \Psi_{0,\pm}^j(x)= \left(
\begin{array}{c}
 \sinh\kappa_j(x+\tau_j)\\
  \pm \sinh\kappa_j(x+\tilde{\tau}_j)
\end{array}
\right)\,,
\end{equation}
$\mathcal{H}^D_0\Psi_{0,\pm}^j(x)=
\mathcal{E}_{0,\pm}(j)\Psi_{0,\pm}^j(x)$, where
$\tilde{\tau}_j=\tau_j-\varphi_j(\kappa_j,\mathcal{C})$ . Here the
first (second) set of the states has to be taken for  the odd
(even) values of the index $j$, cf. (\ref{psij0}). The set of $2n$
functions (\ref{eigenj0D}) form a kernel of the matrix
differential operator $\mathcal{A}_n$ of the order $n$. The
un-normalized bound states of $\mathcal{H}^D_n$ are given by
$\Psi_{n,\pm}^j(x)=\mathcal{A}_n \frac{1}{\kappa_j}\frac{d}{dx}
\Psi_{0,\pm}^j(x)$, cf. (\ref{psitdif}). The normalized bound
states of $\mathcal{H}^D_n$ can be expressed in terms of
eigenstates of the associated supersymmetric Schr\"odinger pair of
the systems, $\hat{\Psi}_{n,\pm}^{jT}(x)=
\frac{1}{\sqrt{2}}(\hat{\psi}_{n,j}(x),\,\pm
\hat{\tilde{\psi}}_{n,j}(x)) $,
$\mathcal{H}^D_n\hat{\Psi}_{n,\pm}^{j}=
\mathcal{E}_{n,\pm}(j)\hat{\Psi}_{n,\pm}^{j}$,
 $\mathcal{E}_{n,\pm}(j)=
\pm\sqrt{\mathcal{C}^2-\kappa_j^2}$, $j=1,\ldots,n$, where
$\hat{\psi}_j(x)$ is given by Eq. (\ref{psihat}), and
$\hat{\tilde{\psi}}_j(x)$ are the corresponding eigenstates of
$\tilde{H}_n$. The spectra for the cases $n=1$ and $n=2$ are
illustrated by Figures \ref{fig2}  and \ref{fig3}.

The nontrivial integral for the Dirac system
$\mathcal{H}^D_n$ is
 $\mathcal{P}_{n}= \mathcal{A}_n
(-i\frac{d}{dx}\mathds{1}) \mathcal{A}_n^\dagger$, which is the
central charge $\mathcal{P}_{n,1}$
of the associated reflectionless
supersymmetric Schr\"odinger system $\mathcal{H}_n$. It is this
operator that distinguishes the degenerate eigenstates
 $\Psi_{n,\pm}^k(x)$ with $k>0$ and $k<0$
 in the continuum part of the spectrum of $\mathcal{H}^D_n$,
$\mathcal{P}_{n}\Psi_{n,\pm}^k(x)=k\prod_{j=1}^n
(\mathcal{E}_{n,\pm}(k)+\kappa_j^2)\Psi_{n,\pm}^k(x)$. It also
detects all the $2(n+1)$ non-degenerate eigenstates of
$\mathcal{H}^D_n$ by annihilating them. The $2n$ of these states
correspond to the bound states inside the energy gap between the
positive and negative continuums in the spectrum of
$\mathcal{H}^D_n$.  The two other states are the states
$\Psi_{n,\pm}^0(x)$ at the edges of the gap.

The trivial Lie-algebraic relation $[\mathcal{H}^D_n,
\mathcal{P}_{n}]=0$ does not show by itself a special nature of
the higher-order matrix integral $ \mathcal{P}_{n}$. This can be
revealed by identification of the own supersymmetric structure of
the Dirac reflectionless system $\mathcal{H}^D_n$. Consider the
following  operator
\begin{equation}\label{defgam}
    \Gamma=R_x\mathcal{R}_{\tau,\mathcal{C}}    \sigma_3\,.
\end{equation}
Here $R_x$ is a reflection operator for variable $x$,
$R_xx=-xR_x$, $R_x^2=1$, whereas $\mathcal{R}_{\tau,\mathcal{C}}$
makes the same job for the parameters $\tau_j$ and $\mathcal{C}$,
$\mathcal{R}_{\tau,\mathcal{C}}\tau_j=-\tau_j
\mathcal{R}_{\tau,\mathcal{C}}  $, $j=1,\ldots, n$,
$\mathcal{R}_{\tau,\mathcal{C}} \mathcal{C}=
-\mathcal{C}\mathcal{R}_{\tau,\mathcal{C}}  $,
$\mathcal{R}_{\tau,\mathcal{C}}^2=1$. Operator (\ref{defgam})
commutes with the Hamiltonian $\mathcal{H}^D_n$  and anticommutes
with $ \mathcal{P}_{n}$, $[\Gamma,\mathcal{H}^D_n]=0$, $\{\Gamma,
\mathcal{P}_{n}\}=0$. Since $\Gamma^2=1$, (\ref{defgam})
can be
treated as the $\Z_2$-grading operator, which identifies
$\mathcal{H}^D_n$ and $ \mathcal{P}_{n}$ as bosonic and fermionic
operators, respectively. So, the reflectionless Dirac system
$\mathcal{H}^D_n$ is described by its own exotic supersymmetry
given by a nonlinear superalgebraic relation
\begin{equation}
    \{\mathcal{P}_{n},\mathcal{P}_{n}\}=
    2\P^D_{2(n+1)}\,,\qquad
    \P^D_{2(n+1)}\equiv \left((\mathcal{H}^D_n)^2-\mathcal{C}^2\right)
    \prod_{j=1}^n\left((\mathcal{H}^D_n)^2+\kappa_j^2
    -\mathcal{C}^2\right)\,.
\end{equation}
The roots of the polynomial
$\P^D_{2(n+1)}(\mathcal{H}^D_n)$
correspond to the $2(n+1)$ non-degenerate eigenvalues of the
Hamiltonian $\mathcal{H}^D_n$. Denoting $\mathcal{P}^D_{n,1}=
\mathcal{P}_{n}$ and defining $\mathcal{P}^D_{n,2}=
i\Gamma\mathcal{P}^D_{n,1}$
as a second supercharge,
a nonlinear $N=2$ superalgebra is generated
for the
$n$-soliton Dirac system:
$\{\mathcal{P}^D_{n,a},\mathcal{P}^D_{n,b}\}=
    2\delta_{ab}\P^D_{2(n+1)}$,
$[\mathcal{H}^D_n,\mathcal{P}^D_{n,a}]=0$, $a,b=1,2$.

It may seem  that the nature of the grading operator
(\ref{defgam}) is rather unusual\footnote{For 
other appearances of exotic supersymmetric structures
based on the  grading operators related to reflections
see \cite{PAN,PlyNi,MPhid,CJP}. }
since it includes in its structure
the operator anticommuting with
 $\mathcal{C}$, that in the case $n=0$ is
 just a mass parameter.
 Recall that
$\mathcal{C}$ can be presented in terms of the parameters $\tau_j$
and $\tilde{\tau}_j$ constrained by the relation (\ref{CCC}), i.e.
$\mathcal{C}=\kappa_j \coth\kappa_j(\tau_j-\tilde{\tau}_j)$,
$j=1,\ldots,n$. Then we see that the operator
$\mathcal{R}_{\tau,\mathcal{C}}$ can alternatively be treated in a
more symmetric way as the operator
$\mathcal{R}_{\tau,\tilde{\tau}}$, which reflects the soliton
translation parameters, $\mathcal{R}_{\tau,\tilde{\tau}}\tau_j=
-\tau_j\mathcal{R}_{\tau,\tilde{\tau}}$,
$\mathcal{R}_{\tau,\tilde{\tau}}\tilde{\tau}_j= -\tilde{\tau}_j
\mathcal{R}_{\tau,\tilde{\tau}}$,
$\mathcal{R}_{\tau,\tilde{\tau}}^2=1$.

\section{Concluding comments and outlook}

We have constructed a quantum reflectionless fermion system,
which corresponds to  the Dirac particle
in a fixed background of a multi-kink-antikink soliton
$\Delta_n(x)$.
The $(2n+1)$-parametric
function $\Delta_n(x)$ can be considered as an `instant
photograph'  of a $2n$-soliton solution
$v(x,t)=\Delta_n(x,t)$
to the  mKdV equation given by Eq. (\ref{Deltanxt}).
Parameter $\mathcal{C}$
corresponds here to the same  nonzero asymptotic,
$\Delta_n(x)\rightarrow \mathcal{C}\neq 0$, as $x\rightarrow
-\infty$  and $x\rightarrow
+\infty$, while other $2n$ parameters,
$\kappa_j$ and $\tau_j$,
 are the scaling and
translation soliton parameters.
As we saw, this mKdV solution can be related
to  two distinct  solutions $u^+=
{U}_n(x+6\mathcal{C}^2t)+\mathcal{C}^2$ and
$u^-=\tilde{U}_n(x+6\mathcal{C}^2t)+\mathcal{C}^2$
of the KdV equation by means
of relations $v^2\pm v_x=u^\pm$.
The second order Schr\"odingier operators
$H^\pm=-\frac{d^2}{dx^2} +u^\pm$
are factorized then in terms of the first order
operators $A=\frac{d}{dx}+v$ and $A^\dagger= -\frac{d}{dx}+v$,
$H^+=AA^\dagger$, $H^-=A^\dagger A$,
which have a sense of the
Darboux intertwining operators, $A^\dagger H^+=H^-A^\dagger$,
$AH^-=H^+A$. In the most  generic case
of real nonsingular potentials $u^+$ and $u^-$,
the first order scalar Darboux
intertwining operators may relate either

{\bf i)} a completely isospectral pair of 1D Schr\"odinger
Hamiltonians, or

{\bf ii)} almost isospectral Hamiltonians
with spectra different only in one bound (ground)
state.

When  nonsingular $u^+$ and $u^-$ are  two distinct finite-gap
solutions to the KdV equation, the first possibility i) may
correspond either to the case of two completely isospectral
finite-gap periodic (or almost periodic) systems, or to a pair of
completely isospectral  $n$-soliton systems. We investigated here
the  soliton case with $A=X_n$, $H^+=H_n+\mathcal{C}^2$ and
$H^-=\tilde{H}_n+\mathcal{C}^2$, which can be considered as an
infinite-period limit of some isospectral pair of finite-gap
periodic systems. The exotic supersymmetric  structure of
isospectral one-gap periodic pairs of the Schr\"odinger (Lam\'e)
systems, and the corresponding Dirac particle in the kink-antikink
crystal were investigated in \cite{PAN} in the context of physics
related to the Gross-Neveu model.
It would be very interesting to
generalize the analysis for the case of periodic finite-gap
systems with the number of prohibited bands $n>1$.

 The second possibility ii) corresponds to
the situation when the quantum systems $H^+$ and $H^-$ are given
by $n$- and $(n-1)$-soliton reflectionless potentials having,
respectively, $n$ and $n-1$ bound states of the same energy,
except the ground state of the $n$-soliton potential having in
this case zero energy. In the simplest case, such a picture is
realized by the pairs of  reflectionless P\"oschl-Teller systems
\cite{CJP}.  The general case of almost
isospectral soliton pairs given by Eqs. (\ref{VVW}) and
(\ref{AAH}) requires a separate consideration.
This, particularly, will give us  a
possibility to relate fermion systems in multi-kink-antikink
backgrounds considered here and characterized by zero topological
number, with fermion systems in the kink-type backgrounds with
nonzero values of a topological charge, and to investigate exotic
supersymmetric structure appearing in the extended Dirac systems.
Such a generalization of the results obtained here seems to be
interesting, particularly, from the perspective of their
application to the physics of carbon nanostructures.

We considered  the quantum mechanics of the Dirac particle in a
fixed background of a multi-kink-antikink soliton.  The
multi-kink-antikink, as well as kink type solitons are also
interesting from another perspective,  related to the physics
associated with the BdG equations \cite{NNB,deG}.

In many physical applications
reflectionless potentials $\Delta(x)$
appear as  stationary
solutions for fermion self-consistent
inhomogeneous
condensates. These are given by
the system of (1+1)D Dirac equations
\begin{equation}\label{SE1}
    \left( i \partial \!\!\!/ - \Delta\right) \psi_{\alpha} =0,
\end{equation}
subject to the constraint
\begin{equation}\label{SE2}
    \Delta=-g^2 \sum_{\alpha=1}^N
    \sum_{\rm occ} \bar{\psi}_{\alpha}\psi_{\alpha}\,,
\end{equation}
where  $\sum_{\alpha=1}^N$ corresponds to summation in
degenerate states, with $\alpha$ denoting
 a generalized flavour
(possibly, including spin) index, and $\sum_{\rm occ}$ is a sum
over the energy levels occupied by each  flavour\footnote{Note
some similarity of (\ref{SE1}) and (\ref{SE2}) with equations
(\ref{psijN+}) and (\ref{UNKayMos}).}. Particularly, these
equations appear in the superconductivity, in the physics of
conducting polymers, and in the Gross-Neveu model
\cite{GN,DHN,polymers,ThiesM,Stone,BarKu,PapaN}.

In the context of the Bardeen-Cooper-Schrieffer theory of
superconductivity, $\Delta$ corresponds to a  `pair potential'. It
is a phonon field generated by moving electrons
 via their interaction with ions.
 Dirac equation  (\ref{SE1})
 with $N=1$ appears in the BdG method
 after diagonilizing the effective mean field
 Hamiltonian by application of
 Bogoliubov transformations,
and by making  use of  the
Andreev approximation, which
 corresponds to linearization of the non-relativistic
 energy dispersion  near the Fermi points, or,
 equivalently, by neglecting second
derivatives of the Bogoliubov
amplitudes \cite{Andr}.
The so called gap
equation, or self-consistency equation (\ref{SE2}) for the pair
potential appears in the theory from a condition of stationarity
of the free energy \cite{Stone}.
 In the physics of conducting polymers,
$\Delta$ corresponds to the order parameter. The order parameter
is related to the Peierls instability, which underlies the
phenomenon of charge and fermion-number fractionalization
\cite{Jackiw,polymers,NieSemSod}. In the Gross-Neveu model
\cite{GN}, being a (1+1)D toy model  for strong interactions that
mimics several  important properties of QCD,  the term
$-g^2\sum_{\alpha=1}^N
  \bar{\psi}_\alpha\psi_\alpha$ corresponds to a nonlinear
  interaction of fermions with $N$ flavors.
  As it was demonstrated by
  Dashen, Hasslacher and Neveu
  \cite{DHN}, in the
     t'Hooft limit $N\rightarrow\infty$, with  $g^2N$ fixed,
 the model can be reduced  to the quasi-classical
 model (\ref{SE1}),
 (\ref{SE2}) \cite{PapaN}. Particularly, they showed that for
 stationary solutions, the Schr\"odinger potentials $V_\pm=
 \Delta^2\pm \Delta_x-\Delta_0^2$ have to be reflectionless.
Their results were developed in diverse directions in
\cite{Feinberg,ThiesM,SchonTh,FeinHill,BasD,PAN,TakNit,DunneThies}.

In the stationary case the Dirac equation (\ref{SE1})
takes the form
 $(\gamma^0E+i\gamma^1\partial_x-\Delta)\psi(x)=0$, where
 we omitted the generalized flavour index $\alpha$.
With the choice $\gamma^0=\sigma_1$ and $\gamma^1=-i\sigma_3$,
this is reduced to the equation $\mathcal{H}_n^D\psi=E\psi$, where
$\mathcal{H}_n^D$ is given by Eq. (\ref{HDir}). Therefore, to
relate the system we studied with the BdG system it is necessary
to provide an appropriate  interpretation for the consistency
equation (\ref{SE2}) by making use of the obtained results. We are
going to present the corresponding investigation elsewhere, having
also in mind a relation between condensates with zero and nonzero
topological charges that has been indicated above.

\vskip0.2cm

\noindent \textbf{Acknowledgements.}
The work of MSP has been partially supported by
 FONDECYT Grant No. 1130017, Chile.
 A. A. is supported
by BECA DOCTORAL CONICYT 21120826.
MSP  is grateful to
Salamanca University, where a part of this work was done,
 for hospitality.

\section*{Appendix A:\,  Darboux transformations}\label{ap1}
\renewcommand{\theequation}{A.\arabic{equation}}
\setcounter{equation}{0}


Here we summarize shortly the basic aspects of  Darboux
transformations used in the main text and in Appendix B.

Let $\psi(x)$ be an eigenstate of the second order Schr\"odinger
operator $H=-\frac{d^2}{dx^2}+u(x)$  of eigenvalue $E$,
$H\psi=E\psi$. Then
\begin{equation}\label{PpsiE}
    u-E=\phi^2+\phi_x\,,\quad
    \text{where}\quad
    \phi\equiv(\ln\psi)_x\,.
 \end{equation}
Define the first order differential operator
$
    A\equiv\psi\frac{d}{dx}\frac{1}{\psi}=\frac{d}{dx}
    -\phi.
$ By definition, $\psi$ is a kernel of $A$, $A\psi=0$, while
$\frac{1}{\psi}$ is a kernel of the Hermitian conjugate operator $
A^\dagger=-\frac{1}{\psi}\frac{d}{dx}\psi=-\frac{d}{dx} -\phi$.
By Eq. (\ref{PpsiE}), shifted for a constant
Hamiltonian $H$ is factorized as
\begin{equation}\label{defH}
    A^\dagger A=H-E\,.
\end{equation}
Define another Hamiltonian operator
$\hat{H}=-\frac{d^2}{dx^2}+\hat{u}$ by
\begin{equation}\label{defHhat}
    AA^\dagger=\hat{H}-E\,,
\end{equation}
so that $\hat{u}-E=\phi^2-\phi_x$. If potential $u(x)$ is
non-singular, and eigenfunction $\psi(x)$ is nodeless, then
$\hat{u}=u-2\phi_x$
 is also a non-singular potential; otherwise it will
have singularities at zeros of $\psi(x)$. Note that the function
$\phi$ can be expressed in terms of the pair of potentials $u$ and
$\hat{u}$ as
\begin{equation}\label{phisuper}
    \phi=\frac{1}{2}\frac{u_x+\hat{u}_x}{u-\hat{u}}.
\end{equation}

In accordance with  (\ref{defH}) and (\ref{defHhat}),
operators $A$ and $A^\dagger$
intertwine the Hamiltonians
$H$ and $\hat{H}$,
$
    AH=\hat{H}A
$,
$
    A^\dagger\hat{H}=HA^\dagger
$. As a consequence, if $\psi_\lambda$ is an eigenstate of $H$ of
eigenvalue $\lambda\neq E$, $H\psi_\lambda=\lambda\psi_\lambda$,
then $A\psi_\lambda \equiv\hat{\psi}_\lambda$ is an eigenstate of
$\hat{H}$ of the same eigenvalue,
$\hat{H}\hat{\psi}_\lambda=\lambda\hat{\psi}_\lambda$. The
operator $A^\dagger$ acts in the opposite direction as
$A^\dagger\hat{\psi}_\lambda=(\lambda -E)\psi_\lambda$.

The described picture corresponds to the Darboux transformation
generated by the first order differential operators $A$ and
$A^\dagger$, which transform mutually the eigenstates of the
Schr\"odinger operators $H$ and $\hat{H}$ with any eigenvalue
$\lambda\neq E$. For eigenvalue $E$, the second, linear
independent from $\psi$ solution of the Shcr\"odinger equation can
be presented as
\begin{equation}\label{secondsol}
    \tilde{\psi}(x)=\psi(x)\int^x\frac{d\xi}{\psi^2(\xi)}\,,
\end{equation}
$W(\psi,\tilde{\psi})=1$.
The action of the $A$ on it produces the
kernel of $A^\dagger$, $
    A\tilde{\psi}\equiv\hat{\tilde{\psi}}=\frac{1}{\psi},
$ $A^\dagger\hat{\tilde{\psi}}=0$.
As a consequence,  $H\tilde{\psi}=(A^\dagger A+E)\tilde{\psi}=E\tilde{\psi}$,
and, on the other hand, $
    \hat{H}\hat{\tilde{\psi}}=(AA^\dagger +E)\hat{\tilde{\psi}}
    =E\hat{\tilde{\psi}}.
$
Analogously, the second eigenstate
of $\hat{H}$ of the eigenvalue $E$ is
\begin{equation}
    \tilde{\hat{\tilde{\psi}}}(x)=\frac{1}{\psi(x)}
    \int^x\psi^2(\xi)d\xi\,.
\end{equation}
The application of $A^\dagger$ to it produces the kernel of $A$,
$A^\dagger\tilde{\hat{\tilde{\psi}}}=-\psi$.

\section*{Appendix B:\,  KdV and mKdV equations,
and Miura
transformation}\label{bp1}
\renewcommand{\theequation}{B.\arabic{equation}}
\setcounter{equation}{0}


Here we describe
shortly the relation between the KdV equation
\begin{equation}\label{KdV}
    u_{t}-6uu_x+u_{xxx}=0,
\end{equation}
and the modified KdV equation (mKdV)
\begin{equation}\label{mKdV}
    v_{t}-6v^2v_x+v_{xxx}=0.
\end{equation}

Given a function  $v=v(x,t)$, let us define another
function $u^+=u^+(x,t)$ by
\begin{equation}\label{udefv}
    u^+=v^2+v_x.
\end{equation}
Assume  that  $v=v(x,t)$ satisfies the mKdV equation (\ref{mKdV}).
Then   $u^+_t=(2v+\partial_x)(6v^2v_x-v_{xxx})$ and
$-6u^+u^+_x+u^+_{xxx}=-(2v+\partial_x)(6v^2v_x-v_{xxx})$, and so
function (\ref{udefv})
defined in terms of some solution of the
mKdV equation satisfies automatically the KdV equation.

The mKdV equation (\ref{mKdV})
 is invariant under  the change $v\rightarrow -v$, while (\ref{udefv})
transforms into
\begin{equation}\label{udefv-}
    u^-=v^2-v_x.
\end{equation}
Therefore, function $u^-$ defined by (\ref{udefv-}) in terms of a
solution of the mKdV equation also satisfies the KdV equation.

Consider now relations  (\ref{udefv}) and (\ref{udefv-}) from
another perspective. Let us assume that we are given a function
$u^+(x,t)$, and treat relation (\ref{udefv}) as a nonlinear
Riccati equation that defines function $v$. If we assume that
$u^+=u^+(x,t)$ satisfies the KdV equation (\ref{KdV}), then we
find that the function $v(x,t)$ defined by  (\ref{udefv})
satisfies not the mKdV, but the equation
\begin{equation}\label{d+mKdV}
    (2v+\partial_x)(v_{t}-6v^2v_x+v_{xxx})=0.
\end{equation}
{}From the latter it follows a relation
$v_{t}-6v^2v_x+v_{xxx}=C(t) \exp(-2\int^xv(\xi,t)d\xi)$, where
$C(t)$ is an arbitrary function. This is reduced to the mKdV
equation only in a particular case of $C(t)=0$. In the described
interpretation, relation (\ref{udefv}) corresponds to the Miura
transformation $u^+\rightarrow v$
\cite{Miura}, which can be
compared with Eq. (\ref{PpsiE}).

If instead of (\ref{udefv}) we define a function $v$ by
(\ref{udefv-}), and assume that
 $u^-(x,t)$ satisfies the KdV equation,
then instead of (\ref{d+mKdV}) we obtain the equation
\begin{equation}\label{d-mKdV}
    (2v-\partial_x)(v_{t}-6v^2v_x+v_{xxx})=0.
\end{equation}
For each of the two Miura transformations, (\ref{udefv}) or
(\ref{udefv-}),  a KdV solution generates a function $v$ which
satisfies not the mKdV equation, but the equation of a more
general form,  (\ref{d+mKdV}) or (\ref{d-mKdV}).

Let us assume now that we have two different functions
$u^+=u^+(x,t)$ and $u^-=u_-(x,t)$ given by   (\ref{udefv}) and
(\ref{udefv-}) in terms of one function $v(x,t)$, and suppose that
both functions $u^+$ and $u^-$ satisfy the KdV equation. In this
case function $v(x,t)$ has to satisfy \emph{simultaneously} the
two equations (\ref{d+mKdV}) and (\ref{d-mKdV}). Adding these
equations, we obtain $4v(v_{t}-6v^2v_x+v_{xxx})=0$, that implies
that $v$ has to satisfy the mKdV equation (\ref{mKdV}). Note that
in this case the solution of the mKdV equation can be  expressed
in terms of solutions $u^+$ and $u^-$ of the KdV equation as
$ v=\frac{1}{2}\frac{u^+_x+u^-_x}{u^+-u^-},
$
cf. (\ref{phisuper}).

 We conclude therefore, that if
two different solutions $u^+$ and $u^-$ of the KdV equation can be
expressed by means of relations (\ref{udefv}) and (\ref{udefv-})
in terms of one function $v$, the latter ought to be a solution of
the mKdV equation.


\end{document}